\documentclass[fleqn,usenatbib]{mnras}

\usepackage[T1]{fontenc}
\usepackage{ae,aecompl}

\usepackage{graphicx}	% Including figure files
\usepackage{amsmath}	% Advanced maths commands
\usepackage{amssymb}	% Extra maths symbols
\usepackage{mathtools}
\newcommand{\fesc}{\ifmmode{f_{\rm esc}}\else{$f_{\rm esc}$}\fi}
\newcommand{\fescs}{\ifmmode{f_{\rm esc}^\star}\else{$f_{\rm esc}^\star$}\fi}
\newcommand{\kms}{\ifmmode{{\;\rm km~s^{-1}}}\else{km~s$^{-1}$}\fi}
\newcommand{\fgas}{\ifmmode{{f_{\rm gas}}}\else{$f_{\rm gas}$}\fi}
\newcommand{\cubecm}{\ifmmode{{\rm cm^{-3}}}\else{cm$^{-3}$}\fi}
\newcommand{\ztwo}{\ifmmode{{\rm [Z_2/H]}}\else{[Z$_2$/H]}\fi}
\newcommand{\zthree}{\ifmmode{{\rm [Z_3/H]}}\else{[Z$_3$/H]}\fi}
\newcommand{\lsim}{\lower0.3em\hbox{$\,\buildrel <\over\sim\,$}}
\newcommand{\gsim}{\lower0.3em\hbox{$\,\buildrel >\over\sim\,$}}

\newcommand{\sfr}{\ifmmode{\textrm{M}_\odot \,\textrm{yr}^{-1} \,\textrm{Mpc}^{-3}}\else{M$_\odot$ yr$^{-1}$ Mpc$^{-3}$}\fi}
\newcommand{\hsfr}{\ifmmode{\textrm{M}_\odot\, \textrm{yr}^{-1}}\else{M$_\odot$ yr$^{-1}$}\fi}

\newcommand{\eavg}{\ifmmode{\langle E_\gamma \rangle}\else{$\langle E_\gamma \rangle$}\fi}

\newcommand{\enzo}{{\sc enzo}}
\newcommand{\yt}{{\sc yt}}

\newcommand{\Ms}{\ifmmode{M_\odot}\else{$M_\odot$}\fi}
\newcommand{\vrms}{\ifmmode{v_{\rm rms}}\else{$v_{\rm rms}$}\fi}

\newcommand{\hh}{H$_2$}

\newcommand{\tvir}{\ifmmode{T_{\rm{vir}}}\else{$T_{\rm{vir}}$}\fi}
\newcommand{\mvir}{\ifmmode{M_{\rm{vir}}}\else{$M_{\rm{vir}}$}\fi}
\newcommand{\rvir}{\ifmmode{r_{\rm{vir}}}\else{$r_{\rm{vir}}$}\fi}

\newcommand{\jj}{\ifmmode{J_{21}}\else{$J_{21}$}\fi}
\newcommand{\flw}{\ifmmode{F_{LW}}\else{$F_{LW}$}\fi}
\newcommand{\kph}{\ifmmode{k_{\rm ph}}\else{$k_{\rm ph}$}\fi}

\newcommand{\msun}{{\rm\,M_\odot}} 

\newcommand{\zsun}{\ifmmode{\rm\,Z_\odot}\else{$\rm\,Z_\odot$}\fi}

\newcommand{\hi}{H {\sc i}}
\newcommand{\hii}{H {\sc ii}}
\newcommand{\hei}{He {\sc i}}
\newcommand{\heii}{He {\sc ii}}
\newcommand{\heiii}{He {\sc iii}}
\newcommand{\nhi}{\ifmmode{N_{\rm HI}}\else{$N_{\rm HI}$}\fi}

\newcommand{\music}{{\sc Music}}
\newcommand{\rockstar}{{\sc Rockstar}}

\def\eps@scaling{1.0}% 
\newcommand\epsscale[1]{\gdef\eps@scaling{#1}}% 
\newcommand\plotone[1]{% 
 \centering 
 \leavevmode 
 \includegraphics[width={\eps@scaling\columnwidth}]{#1}% 
}% 
\newcommand\plottwo[2]{% 
 \centering 
 \includegraphics[width={\eps@scaling\columnwidth}]{#1}% 
 \hfil 
 \includegraphics[width={\eps@scaling\columnwidth}]{#2}% 
}% 

\makeatletter
\newcommand*{\rom}[1]{\expandafter\@slowromancap\romannumeral #1@}
\makeatother

\title[Cradles of the first stars]{Cradles of the first stars: self-shielding, halo masses, and multiplicity}

\author[D. Skinner and J.~H. Wise]{
Danielle Skinner$^{1}$\thanks{E-mail: drenniks@gatech.edu}
and John H. Wise$^{1}$
\\
$^{1}$Center for Relativistic Astrophysics, School of Physics, Georgia Institute of Technology, Atlanta, GA 30332, USA\\
}

\pubyear{2019}

\begin{document}
\label{firstpage}
\pagerange{\pageref{firstpage}--\pageref{lastpage}}
\maketitle

% Abstract of the paper
\begin{abstract}
The formation of Population III (Pop III) stars is a critical step in the evolution of the early universe. To understand how these stars affected their metal-enriched descendants, the details of how, why and where Pop III formation takes place needs to be determined. One of the processes that is assumed to greatly affect the formation of Pop III stars is the presence of a Lyman-Werner (LW) radiation background, that destroys \hh{}, a necessary coolant in the creation of Pop III stars. Self-shielding can alleviate the effect the LW background has on the \hh{} within haloes. In this work, we perform a cosmological simulation to study the birthplaces of Pop III stars, using the adaptive mesh refinement code \textsc{Enzo}. We investigate the distribution of host halo masses and its relationship to the LW background intensity. Compared to previous work, haloes form Pop III stars at much lower masses, up to a factor of a few, due to the inclusion of \hh{} self-shielding. We see no relationship between the LW intensity and host halo mass. Most haloes form multiple Pop III stars, with a median number of four, up to a maximum of 16, at the instance of Pop III formation. Our results suggest that Pop III star formation may be less affected by LW radiation feedback than previously thought and that Pop III multiple systems are common. 
\end{abstract}{}

% Select between one and six entries from the list of approved keywords.
% Don't make up new ones.

%%%%%%%%%%%%%%%%%%%%%%%%%%%%%%%%%%%%%%%%%%%%%%%%%%

%%%%%%%%%%%%%%%%% BODY OF PAPER %%%%%%%%%%%%%%%%%%
%====================================================================
\section{Introduction}
%====================================================================

The first generation of stars transformed a cold and dark universe 
by illuminating and enriching up their neighborhoods with radiation
and heavy elements. The formation of these first stars is a crucial step in the cosmological evolution of the universe because the metals and feedback they deliver to their local environments are necessary for further star production and chemical enrichment. Without these initial stars, heavier metals would not have been produced, and a different universe would be observed than what is observed today. These stars are, by definition, metal-free (Population III; Pop III) and are traditionally thought to be generally massive \citep{ABN02, Bromm02_P3, Turk09, Hosokawa11, Hosokawa16, Hirano15}, although more recent work has shown that lower mass ($M < 1 M_{\odot}$) Pop III stars can be formed \citep{Greif11_P3Cluster, Clark11_Frag, Stacy11}. A substantial fraction of these stars will generate prodigious amounts of ionizing photons and will end their lives in some form of a supernova \citep[e.g.][]{Schaerer02, Heger02}. The supernova will spread the enriched guts of the Pop III star out across its local environment, providing the area with elements the universe has not yet seen. 

Once a halo becomes chemically enriched by the death of Pop III stars, then by definition, it can no longer form more Pop III stars. This marks the end of metal-free star formation in that pregalactic object. Understanding the mixing of metals into the environment of these haloes is necessary to constrain the reach of these metals and the effects they have for future star formation. Numerical studies have shown that turbulence within haloes can mix metals well down to their resolution limit \citep{Wise08_Gal, Greif10, Smith15}. Stars will continue to form in haloes, but with an increased metal abundance. These stars still are metal-poor, having metallicities of 10$^{-6}$ to 10$^{-2}$ of the solar abundance \citep{Chiaki16, Chiaki18, Ritter16}, have a direct chemical connection to Pop III stars, and can survive until the present day \citep{Gnedin06, Tumlinson10, Griffen18, Magg18, Ezzeddine19}. Understanding the formation and chemical abundance of second generation stars can provide more evidence and insight for the earlier generation of Pop III stars.

Without metals and dust to facilitate efficient radiative cooling, Pop III stars rely on \hh{} formation in the gas-phase for cooling. These molecules are however fragile to dissociation from Lyman-Werner (LW) radiation in the energy range 11.2--13.6~eV, through the Solomon process \citep{Field66, Stecher67}. This is a two-step process through which H$_{2}$ is excited to a higher state, H$_{2}^{\ast}$, via absorption of a LW photon:
\begin{equation} \label{Solomon1}
	H_{2} + \gamma \rightarrow  H_{2}^{\ast}
\end{equation}
This excited state then has a probability of dissociating into two hydrogen atoms:
\begin{equation} \label{Solomon2}
	H_{2}^{\ast} \rightarrow H + H
\end{equation}

Diffuse gas is optically thin to LW radiation, thus a background builds over time and can suppress \hh{} formation, delaying Pop III star formation. Furthermore, nearby sources of LW radiation can boost the intensity above the background value which facilitates further \hh{} dissociation in minihaloes. This process can be counteracted with a sufficient amount of \hh{} already present within a halo, via \hh{} self-shielding. Halos with an \hh{} column density of N$_{\rm H2}$ $\geq$ 10$^{14}$ cm$^{-2}$ can suppress the photodissociation of \hh{} by LW photons \citep{Draine96}, giving the halo a chance to form Pop III stars. Other sources of the suppression of Pop III star formation are streaming baryonic velocities \citep{Tselia11, Greif11_Delay, Naoz12, OLeary12, Schauer19, Hirano17_Science}, arising from recombination, and dynamical heating, occurring during the virialization of haloes \citep{Yoshida03, Fernandez14}. X-ray heating from high-mass X-ray binaries also has the potential to suppress Pop III star formation \citep{Jeon14}. 

Given the paradigm of hierarchical structure formation, haloes grow through smooth accretion and successive mergers of smaller haloes. But in which haloes do these first generations of stars form? Their formation rates and locations are important to constrain because they influence the very beginnings of galaxy formation and cosmic reionization. Various semi-analytic investigations have been conducted to learn more about the halo collapse criterion, and thus, which haloes can host Pop III stars. \citet{Tegmark97} discovered that haloes can have a strongly redshift dependent, minimum baryonic mass of 10$^{6}$ M$_{\odot}$ at $z \approx$ 15. In particular, they derive an analytic expression for the fraction of \hh{} needed in a halo for efficient cooling, and determine which  haloes can cool in a Hubble time. \citet{Visbal18} devised a semi-analytic model for the formation of Pop III stars and the transition to metal-enriched stars. They find that varying the Pop III star formation efficiency, the time from a Pop III supernova to metal-enriched star formation, the external enrichment of the halo, and the ionizing escape fraction, leads to large differences in the star formation history of Pop III and metal-enriched stars. This method is useful for exploring the wide parameter space of star formation in the early universe, and could lead to further constraints in the future. 

Previous work have mainly investigated and established the lower limit for a halo to host Pop III stars, or for a halo to collapse. However, subsequent simulations have shown that they do not necessarily form at this minimum due to the aforementioned physical processes playing a role. \citet{Mebane18} used a semi-analytic model of early star formation and found that the LW background coming from the rapidly increasing supply of Pop III stars becomes responsible for suppressing Pop III star formation. \citet{Griffen18} also found that the LW background can significantly suppress the amount of potential sites for Pop III star formation. 

Numerical simulations have also been employed to investigate collapse thresholds and Pop III star formation. \citet[hereafter M01]{Machacek01} found similar results to those mentioned previously, in that the LW feedback can suppress the collapse of small mass haloes. They fit a simple analytic expression for the mass threshold of haloes given a particular LW intensity $J_{\rm LW}$:
\begin{equation} \label{mthresh}
	M_{\rm TH} ( M_{\odot} ) = 1.25 \times 10^{5} + 8.7  \times 10^{5} \left( \frac{4 \pi J_{\rm LW}}{10^{-21}} \right)^{0.47} .
\end{equation}
We will compare our results with this relation in future sections. \citet{Yoshida03} also found that cooling is inefficient in haloes with a LW background > 0.01 $J_{\rm 21}$, where $J_{\rm 21}$ is a specific intensity of 10$^{-21}$ erg s$^{-1}$ cm$^{-2}$ Hz$^{-1}$ sr$^{-1}$, although with sufficient \hh{} shielding, haloes are able to cool in the given LW background radiation. In fact, \citet{Wise07_UVB} found that central shocks drive \hh{} formation, allowing the halo to cool in a LW background intensity of 1 $J_{21}$. They find that \hh{} cooling is always a dominant process, even in large LW background fluxes. \citet{OShea08} also found that Pop III star formation can occur in relatively high LW backgrounds, implying that the LW background may not be a complete indicator of whether or not Pop III stars will form in a given halo.    

In this paper, we focus on the distribution of host halo masses, not just the minimum, and its dependence on redshift and the LW background at the instance of star formation, augmenting results from prior work. We aim to provide further insight into the host haloes of Pop III stars. In \S 2 the methods of the simulation, implementation of star formation, feedback, and \hh{} self-shielding are described. In \S 3 we present the results of our analysis. In \S 4 we discuss the results in more detail and compare with previous work. In \S 5 we conclude our discussion with a summary of our results and the implications therein.

%====================================================================
\section{Methods}
%====================================================================
\subsection{Simulation setup}
%====================================================================
We run and analyze a cosmological simulation with the adaptive mesh refinement (AMR) code \textsc{Enzo} \citep{Enzo} and the toolkit \yt{} \citep{yt_full_paper}. \textsc{Enzo} uses an N-body adaptive particle-mesh solver \citep{Efstathiou85, Couchman91} to follow the dark matter (DM) dynamics. We use the nine-species (\hi, \hii, \hei, \heii, \heiii, e$^{-}$, H$_{2}$, H$_{2}^{+}$, H$^{-}$) non-equilibrium chemistry model \citep{Abel97, Anninos97}. This simulation is similar to the RP simulation in \citet[hereafter W12]{Wise12_RP}, but with updated cosmological and Pop III parameters, and the inclusion of \hh{} shielding.

We simulate a 1 Mpc$^{3}$ comoving box with a 256$^{3}$ base grid 
resolution and a dark matter particle mass of 2001 M$_{\odot}$. It has a maximum refinement level of 12 which provides a maximal comoving resolution of $\sim$1 pc.  We refine the cells when one of the following criteria are met: (1) the baryon or dark matter overdensity exceeds $3\bar{\rho}_{\rm b,DM} \times 2^{\ell(3+\epsilon)}$, or (2) the local Jeans length is less than eight cell widths.  Here $\rho_{\rm b,DM}$ is the baryon or dark matter cosmic mean density, $\ell$ is the AMR level, and $\epsilon = -0.2$ invokes super-Lagrangian refinement behavior. The simulation is initialized with \music{} \citep{Hahn11_MUSIC} at $z = 130$ and uses the cosmological parameters from the Planck collaboration best fit \citet{Planck13_Cosmo}: $\Omega_{\rm M}$ = 0.3175, $\Omega_{\Lambda}$ = 0.6825, $\Omega_{\rm DM}$ = 0.2685, and $h = 0.6711$, with the variables having their usual definitions. We note that this initialization redshift is too low, even with 2nd order Lagrangian perturbation theory, for a small cosmological volume and results in delayed structure formation at very high redshift $z > 20$. The simulation is run until $z = 9.32$, when it becomes too computationally expensive to continue. At this point, about 20\% of the volume is over 10\% ionized. We output 918 datasets,  roughly 0.5 Myr apart. In this paper, we focus only on outputs where Pop III stars have just formed, from $z$ = 27.23 to 9.39. Throughout the rest of the paper, all numbers are in physical units, unless otherwise specified.

A time-dependent LW optically thin radiation background is applied in the simulation. This was fit in W12 (see their Eq. 16) and is consistent with the values in \citet{Trenti09_SFR}. The background evolution of the specific intensity takes the following form:
\begin{equation} \label{LWbg}
	\log_{10}(J_{\rm LW}/J_{21}) = A + Bz - Cz^{2} + Dz^{3} - Ez^{4}
\end{equation}
where $(A, B, C, D, E)$ = $(-2.567, 0.4562, - 0.02680, 5.882 \times 10^{-4}, - 5.056 \times 10^{-6})$.  This form only varies by a factor of five in $z = 9-25$ and has a maximum of $J_{\rm LW}/J_{21} = 0.97$ at a redshift of 13.5 and slowly declines to 0.60 $J_{21}$ at $z = 9.3$ when we halt the simulation. We use adaptive ray tracing \citep{Abel02_RT, Wise11_Moray} to evolve the ionizing radiation field. Centered on each metal-enriched and Pop III star particle, we model the \hh{} dissociating radiation field with an optically thin, inverse square profile. These LW sources are added on top of the background intensity described above.
%====================================================================
\subsection{Star formation}
%====================================================================
In this section, we briefly discuss our implementation of Pop III and metal-enriched star formation. We do not consider the formation and feedback from stellar remnants or asymptotic giant branch (AGB) stars. For further details, we refer the reader to W12. 

%====================================================================
\subsubsection{Pop III Star Formation and Feedback }
%====================================================================
The utilized Pop III star formation model is the same as in \citet{Wise08_Gal} but with updated parameters. Each star particle represents a single massive star, and is formed in a cell when the following criteria are met: 
\begin{enumerate}
	\item a metallicity Z $\leq$ 5 $\times$ 10$^{-6}$ Z$_{\odot}$

	\item a gas number density n > 10$^{6}$ cm$^{-3}$

	\item converging gas flow, $\nabla \cdot \mathbf{v_{gas}}$ < 0 

	\item molecular hydrogen number fraction, f$_{\rm H2}$ > 10$^{-3}$
\end{enumerate}
The critical metallicity marks where dust cooling becomes efficient enough to cause fragmentation at high densities \citep[e.g.][]{Schneider06_Frag}.  In practice, we find about 10 per cent of Pop III star particles having a non-zero metallicity below this value.  This scenario occurs when a halo is externally enriched by a nearby supernova.  Because of uncertainties with turbulent mixing and collapse timescales, we choose not to predict the final metallicity at zero-age main sequence and assign the star particle the metallicity of the densest cell.  The chosen threshold density are consistent with values found in previous simulations of Pop III star formation \citep[e.g.][]{Hirano15} at our resolution limit.  Additionally, it is similar to a density that would trigger another mesh refinement past level 12.  At these scales, Jeans length refinement is the most common because of the cold temperatures in the cloud core.  The molecular hydrogen criterion is also consistent with previous work in the ``loitering'' phase of molecule formation between $10^2$ and $10^8 \cubecm$ before three-body reactions become important \citep[e.g.][]{Omukai10}.

If within 10 pc, multiple cells meet this criteria, then a single Pop III star forms at the center of mass of these cells. In a parameter study, we have found that the number of Pop III star particles is insensitive to this algorithmic merging distance. We ran another 1~Mpc$^3$ simulation with the same setup as the main simulation with the exception that it was a zoom-in simulation focusing on a single minihalo of mass $5 \times 10^5~\Ms$ collapsing at $z \simeq 23$.  Starting from 1~Myr before the collapse, we followed Pop III star formation in this one halo with a merging distance of 10, 5, 2.5, 1.25, and 0.625~pc.  We found that even when this number was decreased to 0.625 pc, the number of star particles that formed was unchanged.  We note that our density threshold for star particle formation is equivalent to a molecular core.  Fragmentation of metal-free gas into massive stellar precursors can occur at scales down to 1000~AU within molecular cores \citep[e.g.][]{Turk09, Stacy11}, therefore our results on multiplicity should be considered as lower limits.

The mass of the Pop III star is randomly sampled from an initial mass function (IMF) of the form:
\begin{equation} \label{IMF}
	f(\log M)dM = M^{-1.3} \exp \left[-\left( \frac{M_{\rm char}}{M}\right)^{1.6} \right]dM
\end{equation}
where M$_{\rm char}$ = 20 M$_{\odot}$ \citep{Hirano17}. This characteristic mass is different from the W12 choice of 100 M$_{\odot}$. This results in an exponential cutoff below M$_{\rm char}$ and a power-law IMF above. The Pop III IMF mass ranges from $1 \leq M / M_{\odot} \leq 300$. Once the cell meets the above criteria and the stellar mass is chosen, an equal amount of gas is removed from the grid in a sphere which contains twice the stellar mass that is centered on the new star particle. The new Pop III star then gains the mass-weighted velocity of the gas contained within that sphere. In this simulation, we do not track any small scale fragmentation which may form low mass stars \citep{Greif11_P3Cluster}. Again, we are only tracking the formation of metal-free molecular cores, which form either a single or multiple massive Pop III stars and perhaps several more low mass stars.

We use the mass-dependent hydrogen ionizing and LW luminosities and lifetimes from \citet{Schaerer02} and a mass-independent photon energy of E$_{\rm ph}$ = 29.6 eV, which is appropriate for the nearly mass-independent surface temperature of Pop III stars, at 10$^{5}$ K. A Type II supernova results if the Pop III star dies with a mass between 11 $\leq$ M$_{\star}$/M$_{\odot}$ $\leq$ 40. A pair instability supernova (PISNe) results if the star dies with a mass between 140 $\leq$ M$_{\star}$/M$_{\odot}$ $\leq$ 260. Outside of these mass ranges, the Pop III star dies as a black hole. In this simulation, no black hole physics are implemented, so the Pop III star simply turns into an inert collisionless particle. When 11 $\leq$ M$_{\star}$/M$_{\odot}$ < 20, the star dies as a normal Type II supernova with an energy of 10$^{51}$ erg. From 20 $\leq$ M$_{\star}$/M$_{\odot}$ $\leq$ 40, the star dies as a hypernova with energies taken from \citet{Nomoto06}. The PISNe have explosion energies from \citet{2002ApJ...567..532H}, of which the analytic function fit to their models can be seen in Eq. 12 of W12. With our chosen IMF (Eq. \ref{IMF}), 59, 36, and 5 per cent of Pop III stars produce Type II SN, black holes with no SN, and PISNe, respectively.

%====================================================================
\subsubsection{Pop II Star Formation and Feedback}
%====================================================================
The Pop II star formation model is the same as \citet{Wise09}, which is equivalent to the above criteria for Pop III star formation but without the molecular hydrogen fraction requirement, and with a metallicity greater than the above value. Star formation is only allowed for gas with temperatures $T$ < 1000 K, to ensure that the volume is cold and collapsing. Contrary to Pop III stars representing single star particles, metal-enriched star particles are modelled as clusters of stars. We enforce a minimum mass of M$_{\rm min}$ = 1000 M$_{\odot}$ for these star particles. 

The lifetime of these particles is 20 Myr, during which they radiate 6000 hydrogen ionising photons per stellar baryon, or 1.13 $\times$ 10$^{46}$ photons s$^{-1}$ M$_{\odot}^{-1}$, which is appropriate for a Salpeter IMF \citep{Schaerer03}. Their spectra are approximated with a monochromatic spectrum with an energy of 21.6 eV at a constant luminosity. After living for 4 Myr, the stars begin to return supernova energies of 6.8 $\times$ 10$^{48}$ erg s$^{-1}$ M$_{\odot}^{-1}$ back to their surroundings.

%====================================================================
\subsection{\texorpdfstring{H$_2$}{H2} Self-Shielding}
%====================================================================
We use the Sobolev-like approximation from \citet{Wolcott11} to model \hh{} self-shielding. To determine the \hh{} shielding factor, the column density of \hh{} needs to be calculated:
\begin{equation} \label{Nh2}
	N_{\rm H2} = n_{\rm H2}L_{\rm char} \ ,
\end{equation}
where n$_{\rm H2}$ is the number density of \hh{} and L$_{\rm char}$ is a characteristic length over which n$_{\rm H2}$ is assumed to be constant. The method employed in this simulation is to define the characteristic length as:
\begin{equation} \label{Lchar}
	L_{\rm char} = \frac{\rho}{|\nabla \rho|} \ . 
\end{equation}
This Sobolev-like method determines the length over which gas with density $\rho$ is diminished. They found that the most accurate, non-ray tracing, method is to use a single Sobolev-like length determined from the mean Sobolev-like length over all Cartesian directions. 

Upon numerically calculating the shielding factor of \hh{} for simulated protogalaxies, Wolcott-Green et al. determined that a slight adjustment to the shielding factor from \citet{Draine96} is sufficient to account for inaccuracies at higher temperatures. The implemented shielding factor is then (see Eq. 10 in \citet{Wolcott11}):
\begin{equation} \label{shield}
	\begin{multlined}
	f_{\rm sh}(N_{\rm H2}, T) = \frac{0.965}{(1+x/b_{5})^{1.1}} + \frac{0.035}{(1+x)^{0.5}}  \times \\ \exp [-8.5 \times 10^{-4} (1+x)^{0.5}]
	\end{multlined}
\end{equation}
Here, $x \equiv N_{\rm H2}/5 \times 10^{14}$ cm$^{-2}$, $b_{5} \equiv b/10^{5}$ cm s$^{-1}$, and $b$ is the Doppler broadening parameter. This shielding factor acts as a multiplier to the \hh{} dissociation rate, where if the shielding factor equals one, there is no shielding, and zero if there is maximum shielding. \citet{Hartwig15} use the same shielding factor equation from \citet{Wolcott11} but implement an improved calculation of the column density to incorporate the relative gas velocities and geometries of the halo. This method takes into account the Doppler-shifting of spectral lines, which can affect \hh{} shielding. With the improved calculation of the column density, they find that the shielding factor is larger for gas at higher temperatures near $10^4$~K than when the shielding factor is calculated for an average column density for the halo, meaning that self-shielding is generally less efficient for higher temperatures. In this work, we do not take into account the Doppler-shifting of spectral lines, but as will be seen below, our haloes are less massive and contain gas at lower temperatures. Our calculation of the shielding factor is still valid for the temperature range of our haloes, and therefore, the Doppler-shifting of spectral lines is not expected to affect our results.

%====================================================================
\subsection{Analytic Calculation of the Shielding Factor for a Static Halo}
%====================================================================
To demonstrate how the shielding factor changes as a function of halo mass and redshift, we can examine $N_{\rm H2}$ between the core and halo virial radius as a function of halo mass for an isothermal halo, with a realistic \hh{} halo profile, and use Equation \ref{shield} to calculate the shielding factor. The \hh{} molecule number density within the halo is given by 
\begin{equation} \label{numberh2}
	n_{\rm H2}(r) = f_{\rm H2} \frac{\rho_{0}R^{2}_{\rm vir}}{\mu m_{\rm H} r^{2}}
\end{equation}
where $\mu = 1.22$ is the mean molecular weight, $\rho_{0} = 200 \rho_{c} / 3$ is the density at the virial radius, and $\rho_{c} = 3 \rm H_{0}^{2} / 8 \pi G (1+z)^3$ is the critical density. Here, H$_{0}$ is the Hubble constant, G is the gravitational constant, and z is the redshift. The column density is then equation \ref{numberh2} integrated from the radius of the core to the virial radius, where the core has a radius of $f_{c} R_{\rm vir}$: 
\begin{equation} \label{columnh2}
	N_{\rm H2} = f_{\rm H2} R_{\rm vir} \frac{200 \rho_{c}}{3 \mu m_{\rm H}} \left(\frac{1}{f_{c}} - 1 \right)
\end{equation}

\begin{figure}
	\includegraphics[width=\columnwidth]{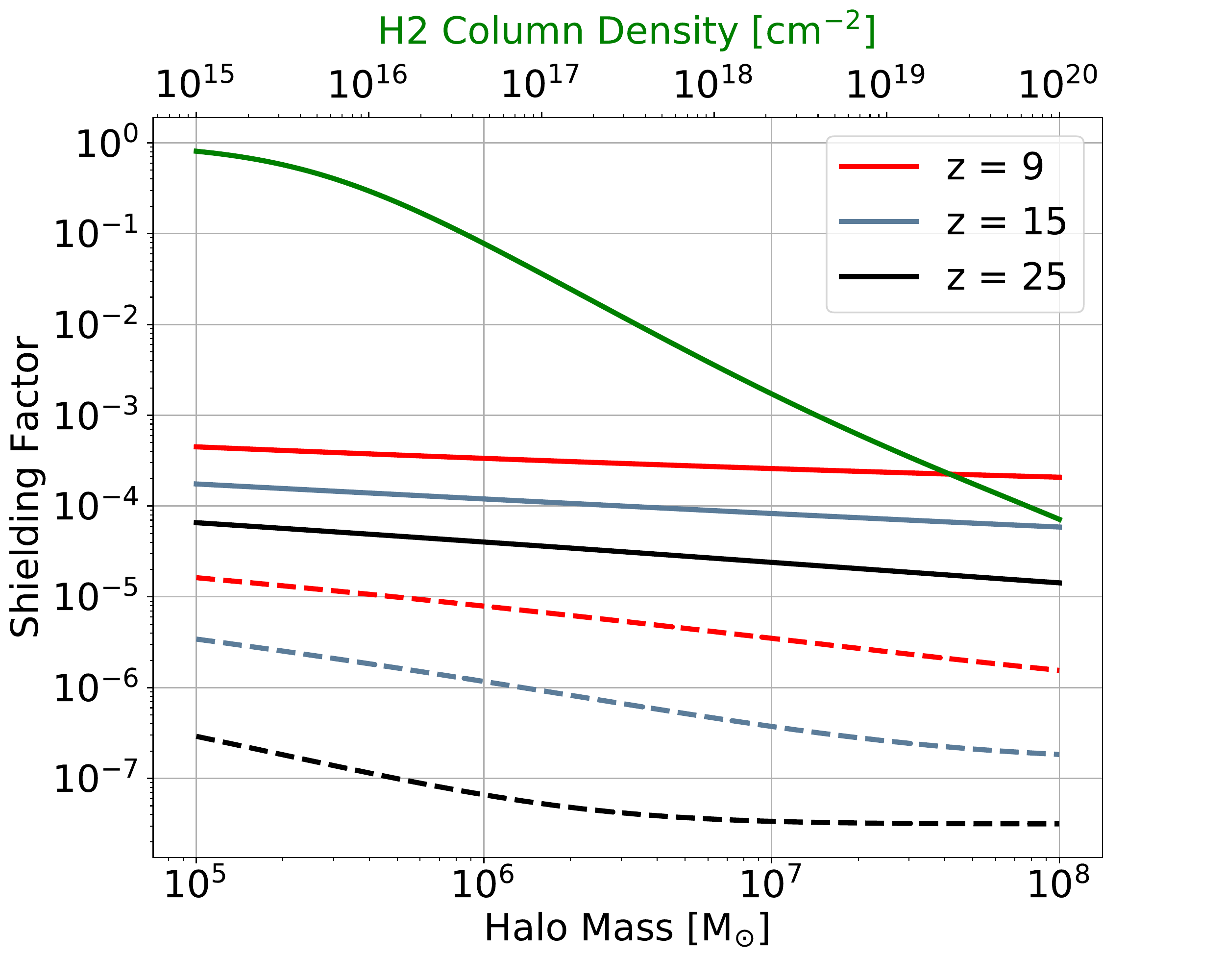}
    \caption{Shielding factor as a function of halo mass. Solid lines indicate a core radius of $f_{c} = 10^{-2}$ and the dashed lines indicate a core radius of $f_{c} = 10^{-4}$. For these parameters, assuming the Doppler broadening parameter $b$ equals the circular velocity, and assuming \hh{} fraction halo profiles from \citet{oshea07a}, the shielding factor is below 1 percent. The green line shows the shielding factor as a function of \hh{} column density for $b = 10 \kms$. Note that a shielding factor of 1 (0) corresponds to no (complete) shielding.}
    \label{fig:shield_mass}
\end{figure}

For $\rm{H}_{0} = 70 \ \rm{km} \ \rm{s}^{-1} \ \rm{Mpc}^{-1}$, and assuming the Doppler broadening parameter, $b$, is equal to the circular velocity of the halo, Equation \ref{columnh2} and Equation \ref{shield} will give the shielding factor for a given core radius, $f_{c}$, redshift, and for a given \hh{} fraction halo profile. Figure \ref{fig:shield_mass} shows the shielding factor as a function of halo mass, for redshifts $z = 9$, $15$, and $25$ in red, blue and black, respectively. The solid lines indicate a core radius of $f_{c} = 10^{-2}$ and the dashed lines a core radius of $f_{c} = 10^{-4}$. To have an accurate depiction of the shielding factor for a given halo radius, the \hh{} fraction is given by the \hh{} fraction halo profiles at $z = 24.1$ from \citet{OShea08} (their Figure 13d) with $J_{21} = 0$, which was extracted directly from their plot. These profiles were approximated with a piecewise power-law for simplicity given by $\log_{10}(f_{\rm H2}) = -0.192 \times \log_{10}(r) - 3.039$ for $r < 3.5$ pc and $\log_{10}(f_{\rm H2}) = -0.745 \times \log_{10}(r) - 2.737$ for $r > 3.5$ pc. 
As the halo masses increase, the shielding factor decreases, meaning there is more shielding occurring within that halo. This behavior occurs because as halo masses increase, more \hh{} is being added to the calculation of the column density. The more \hh{} present within a halo, the smaller the shielding factor will become. Across the top axis and plotted as a green line, the shielding factor as a function of \hh{} column density is shown for a Doppler broadening parameter $ b = 10 \kms$. At each redshift, all halo masses, and for each core radius, we see the shielding factor is significantly smaller than one, implying that \hh{} shielding for these haloes is almost at a maximum. Self-shielding will then significantly decrease the dissociation rate of \hh{} in the halo core, allowing for the presence of a significant \hh{} fraction even in a strong LW background.

%====================================================================
\section{Results}
%====================================================================
In this section, we will present the distribution of host halo masses of Pop III stars, the relation between the mass distribution and the LW background radiation, and the distribution of various Pop III properties.

\begin{figure}
	\includegraphics[width=\columnwidth]{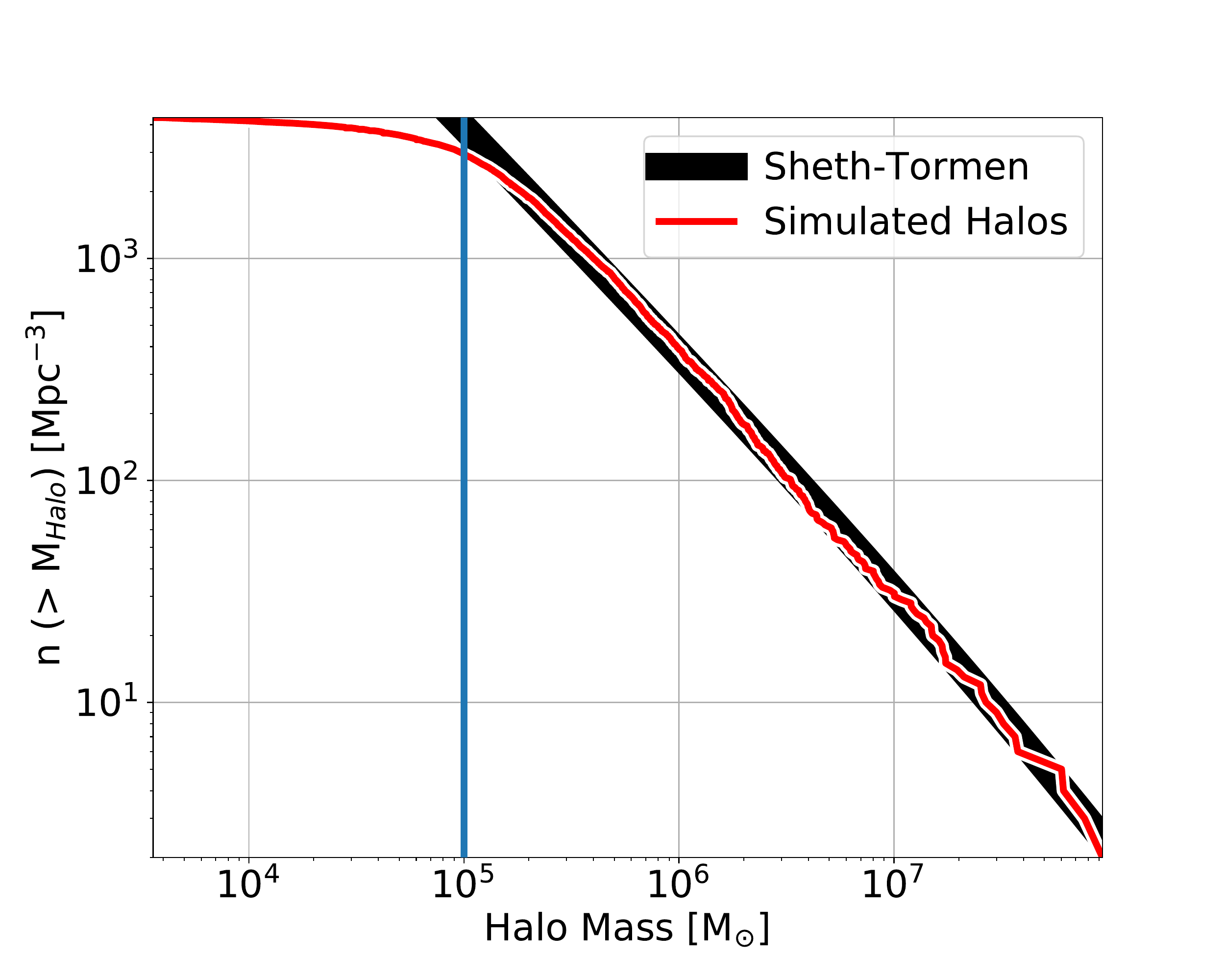}
    \caption{Halo mass function of the last output of the simulation at $z$ = 9.3. The thick, black line is the analytic Sheth-Tormen mass function. The distribution of halo masses matches well with analytical expectations above 10$^{5}$ M$_{\odot}$, where haloes are captured by 50 particles. The vertical blue line indicates our halo mass resolution limit of 10$^{5}$ M$_{\odot}$.}
    \label{fig:hmf}
\end{figure}

\begin{figure}
	\includegraphics[width=\columnwidth]{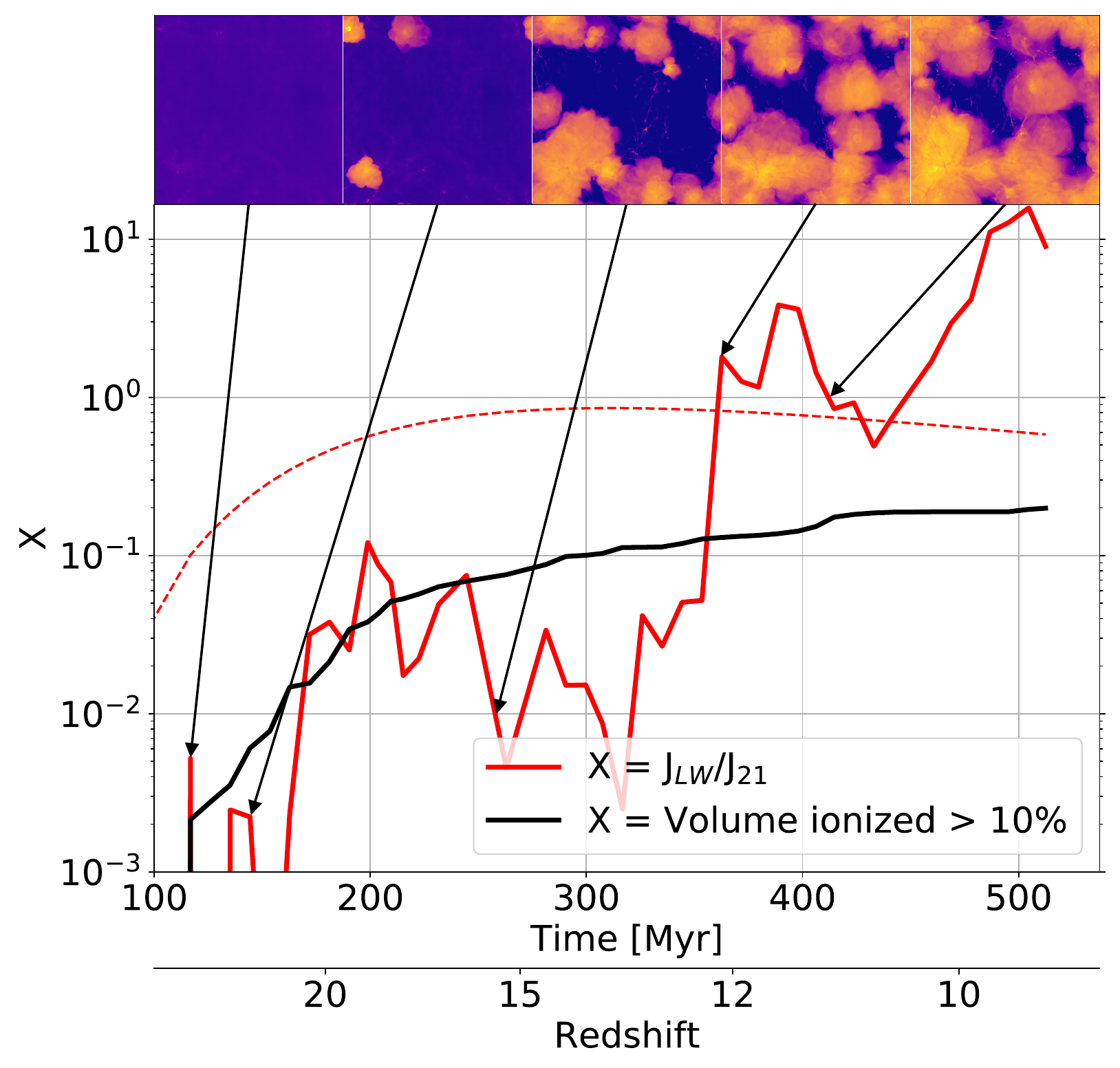}
    \caption{For the entire box, the mass-weighted average LW intensity from stars within the simulation is shown in solid red, and the fraction of the volume ionized above 10 per cent versus time is shown in solid black. The red dashed line represents the LW background implemented in the simulation (see Equation \ref{LWbg}). The solid red line in the legend refers to both the solid and dashed lines in the figure. Temperature projections are shown at the points indicated by the arrows.}
    \label{fig:JLW_xe_mass}
\end{figure}

To ensure that our box is representative of a typical piece of the universe, Figure \ref{fig:hmf} shows the simulated halo mass function along with the analytic Sheth-Tormen mass function at $z = 9.3$. The analytic function closely matches our simulation until the mass resolution of our simulation is unable to resolve below 10$^{5}$ M$_{\odot}$ haloes that contain $\approx$ 50 particles. After a sufficient amount of time after the beginning of the simulation, enough matter has collapsed to begin forming Pop III stars throughout the box. When the first Pop III star forms in our simulation at $z = 27$, there are only 6 haloes that are above $10^5~\Ms$, which are resolved by 50 DM particles, consistent with analytical expectations from an ellipsoidal variant of Press-Schechter formalism \citep{PS74,Sheth01}. These Pop III stars either become black holes or supernovae at the end of their lives, with the latter resulting in an expulsion of metals into the star's surroundings. Metal-enriched star formation begins to take place as soon as enough metals are present within a halo. In the meantime, haloes form and merge with one another, resulting in a wide variety of galaxies, and reionization starts to take place. By the end of the simulation, 20\% of the box has an ionized fraction greater than 10 per cent, as can be seen in Figure \ref{fig:JLW_xe_mass}, which also shows the mass weighted LW intensity coming from stars within the simulation as a function of time, the LW background given by Equation \ref{LWbg} in units of $J_{21}$, and temperature projections of the box at certain points in time. 

\begin{figure}
	\includegraphics[width=\columnwidth]{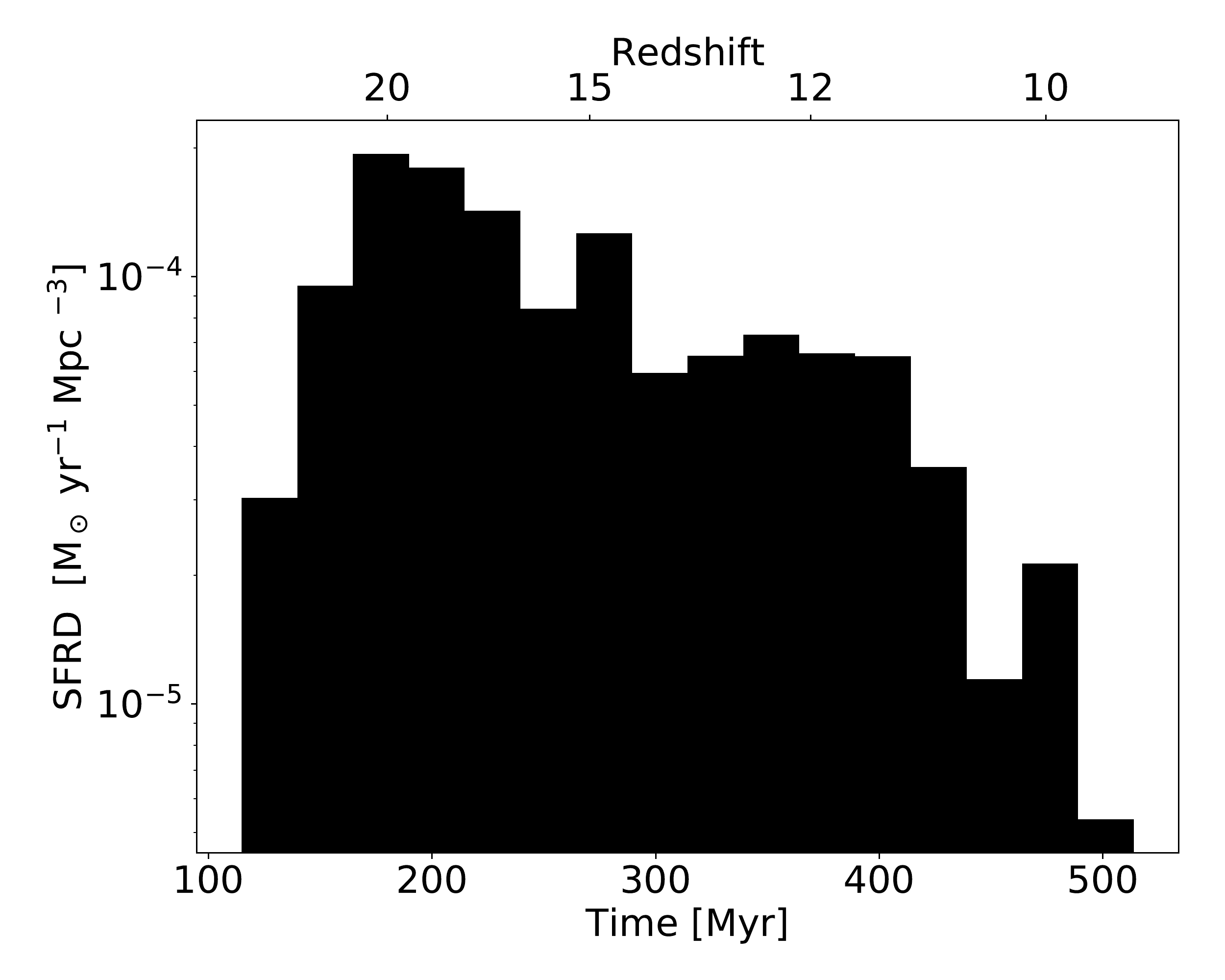}
    \caption{The comoving star formation rate density of Pop III stars. The SFRD peaks at a redshift of 20, and slowly falls off. Note that metal-enriched star formation continually increases throughout time, although that is not plotted here. Due to the small box size of our simulation, this SFRD is particular for our box.}
    \label{fig:pop3_SFR_bar}
\end{figure}

In our simulation, 688 Pop III stars form between 27.23 $\geq z \geq$ 9.39. The corresponding comoving star formation rate density (SFRD) can be seen in Figure \ref{fig:pop3_SFR_bar}, peaking at $1.9 \times 10^{-4} \ \rm M_{\odot} \ yr^{-1} \ Mpc^{-3}$ at $z = 21.3$ and slowly decreases. The SFRD sharply declines near the end of the simulation but does not drop to zero. Small volumes are subject to cosmic variance, in particular the formation time of the first galaxies, which generates the bulk of the LW radiation background.  This SFRD is therefore not representative of a cosmological average.  In particular, it decreases at $z \le 11$ when the most massive galaxy undergoes an extended period of star formation, boosting the local LW radiation field as shown in Figure \ref{fig:JLW_xe_mass}. \citet{Xu13} and \citet{Magg16} see a SFRD of $~10^{-4}~\rm M_{\odot} \ yr^{-1} \ Mpc^{-3}$ at $z = 15$ and $z = 20$ respectively. \citet{Jaacks19} find a SFRD of $~10^{4.5}~\rm M_{\odot} \ yr^{-1} \ Mpc^{-3}$ at $z = 20$, which flattens out to $10^{-3}~\rm M_{\odot} \ yr^{-1} \ Mpc^{-3}$ by $z =7$. Our SFRD rates are consistent with these other studies, but it is likely to change depending on the particular modes captured by the initial conditions.

%====================================================================
\subsection{Population III host halo masses}
%====================================================================
 To determine the host haloes of new Pop III stars, we use the halo finding code, \rockstar{} \citep{rockstar}, an algorithm that adaptively refines haloes found with a friends-of-friends method in position and velocity phase-space and also temporally. The mass determined by \rockstar{} is the dark matter mass, so to determine the total mass of the haloes, we multiply the masses by $\Omega_{\rm M}/\Omega_{\rm DM}$. The total mass is subsequently used in our analysis. For stars that form in subhaloes identified by \rockstar{}, we assign the parent halo mass as the host halo mass.  There is one interesting case that we exclude from our analysis, where the parent halo is above the atomic cooling limit, because the radiative cooling physics are different in this regime and is not sensitive to a LW background.  Here seven Pop III stars form in a pristine $10^6 \msun$ subhalo inside of a $3 \times 10^8 \msun$ atomic cooling parent halo.  Because this object only contains 1 per cent of the number of Pop III stars, it only affects our average halo masses by a similar amount. We also take the center and virial radius from the \rockstar{} halo catalog. About 1\% of Pop III stars do not form in a halo identified by \rockstar{}. Assuming that Pop III stars must form in collapsed haloes, we calculate the virial mass and radius of a halo centered on the Pop III star directly from the simulation data by determining the sphere that contains an average overdensity of 200 $\rho_{\rm c}$. We calculate the LW intensity impinging on each halo by summing up the contribution coming from each radiating star particle outside the virial radius of the host halo. This is then added to the constant LW background implemented in the simulation (Eq. \ref{LWbg}).

\begin{figure}
	\includegraphics[width=\columnwidth]{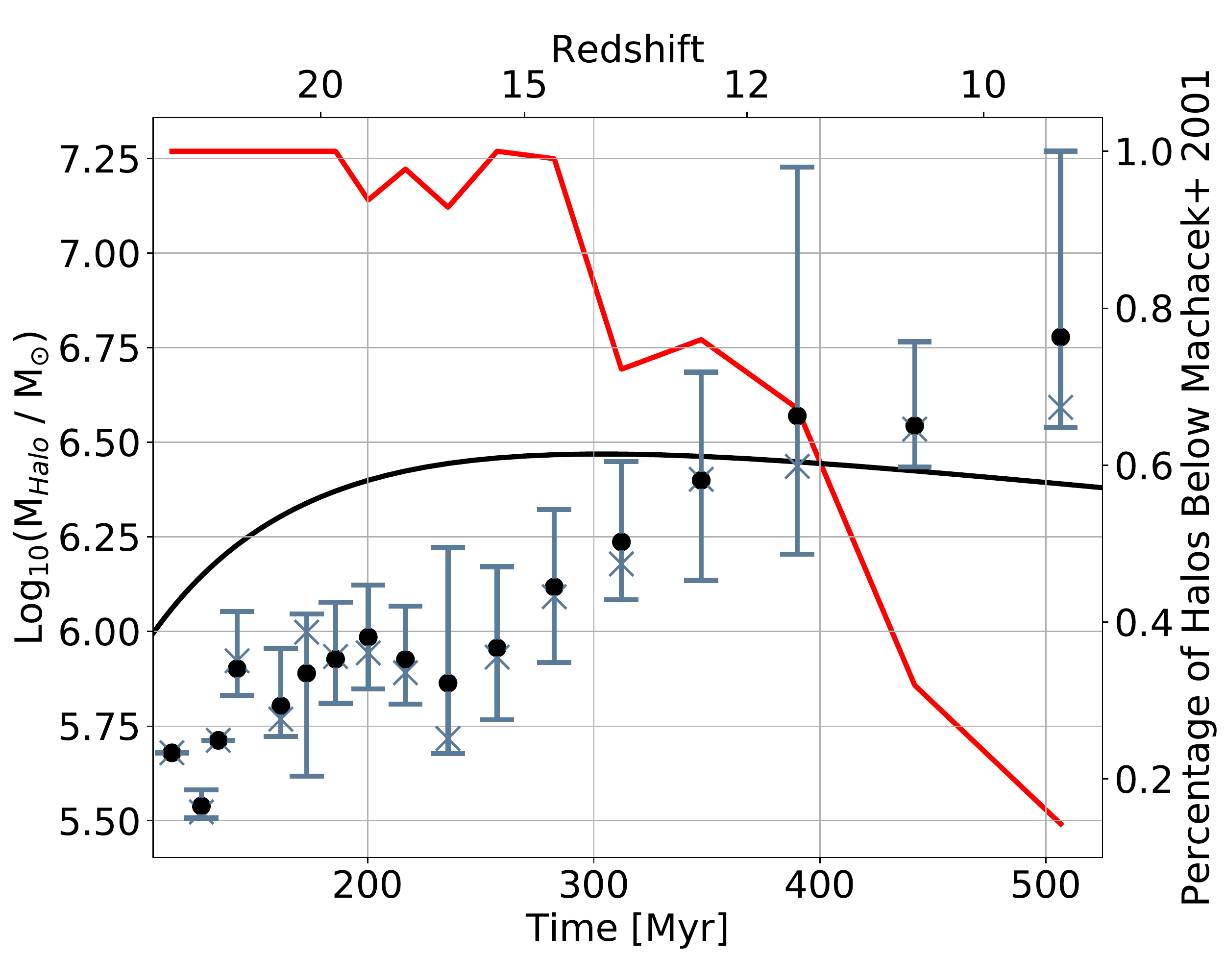}
    \caption{For haloes hosting new Pop III stars, the mean halo mass for redshift bins of $\Delta z$ = 1 is plotted along the left hand side versus time. The black scatter points indicate the mean halo mass for that redshift bin and the x indicate the median halo mass. The error bars indicate the 15.9 and 84.1 percentiles. The right hand side shows the percentage of haloes that fall below the M01 mass threshold as a function of redshift. The M01 mass threshold (Eq. \ref{mthresh}), is plotted for the constant LW background in Eq. \ref{LWbg}. Almost all haloes fall below the M01 relation, until $\approx$ 380 Myr, when the mean halo mass rises above the relation.}
    \label{fig:mean_mass}
\end{figure}

In redshift bins of $\Delta z$ = 1, the mean halo mass hosting Pop III stars is determined, and plotted as black dots against the redshift bins in Figure \ref{fig:mean_mass}. The median for each bin is represented by a cross, and the 15.9 and 84.1 percentiles are plotted below and above the mean, respectively. The M01 relation (Equation \ref{mthresh}) for our LW background intensity in Eq. \ref{LWbg} is shown as the black solid line. Notably, the mean halo mass falls well below the relation, at $<10^{6}$ M$_{\odot}$ until $z = 15$ (270 Myr), at which point, the mean halo mass steadily rises above $3 \times 10^{6}$ M$_{\odot}$. The large discrepancy between our data and the mass threshold from M01 is indicative of the \hh{} shielding which is included in our simulation. \hh{} shielding allows for haloes to cool at much lower masses, and therefore, Pop III stars are forming in these haloes at earlier times. In the M01 analysis, \hh{} shielding is neglected in their calculations for a variety of reasons, including the Doppler broadening of LW bands and large column densities of \hh{} only beginning to form at late times. Interestingly, a full radiation hydrodynamics simulation, such as ours, does not produce identical results as M01 predicts. 

Throughout the literature, there appears to be a consensus that \hh{} self-shielding will help smaller mass haloes collapse in the presence of a LW background radiation \citep[e.g.][]{Yoshida03, Ricotti01, Glover01, Hartwig15}. As can be seen by the red solid line in Figure \ref{fig:mean_mass}, nearly 100\% of the haloes hosting Pop III stars fall below the M01 relation, until around a redshift of 15, at which point, the mean halo mass begins to rise above the relation. The increase in the mean halo mass above the M01 relation is indicative of the increased LW intensity permeating the box due to a galaxy that becomes very active in star formation at late times. This galaxy has a halo mass of $3.6 \times 10^{8} \rm M_{\odot}$, a total stellar mass of $3.1 \times 10^{6} \rm M_{\odot}$, and has a peak Pop III SFR of $2.7 \times 10^{-4}\, \rm M_{\odot}\ yr^{-1}$ at $z=15$ (250 Myr) and a peak metal enriched SFR of $5.4 \times 10^{-2}\, \rm M_{\odot} \ yr^{-1}$ at $z=9.5$ (500~Myr). This galaxy produces a large amount of LW radiation which dominates the box, and drives up the minimum halo mass required for the formation of Pop III stars. This increase in the LW background can also be seen in Figure \ref{fig:JLW_xe_mass}, where by the end of the simulation, the LW background rises to greater than 1 $J_{21}$.

%====================================================================
\subsection{Lyman-Werner intensities in Pop III host haloes}
%====================================================================

\begin{figure}
	\includegraphics[width=\columnwidth]{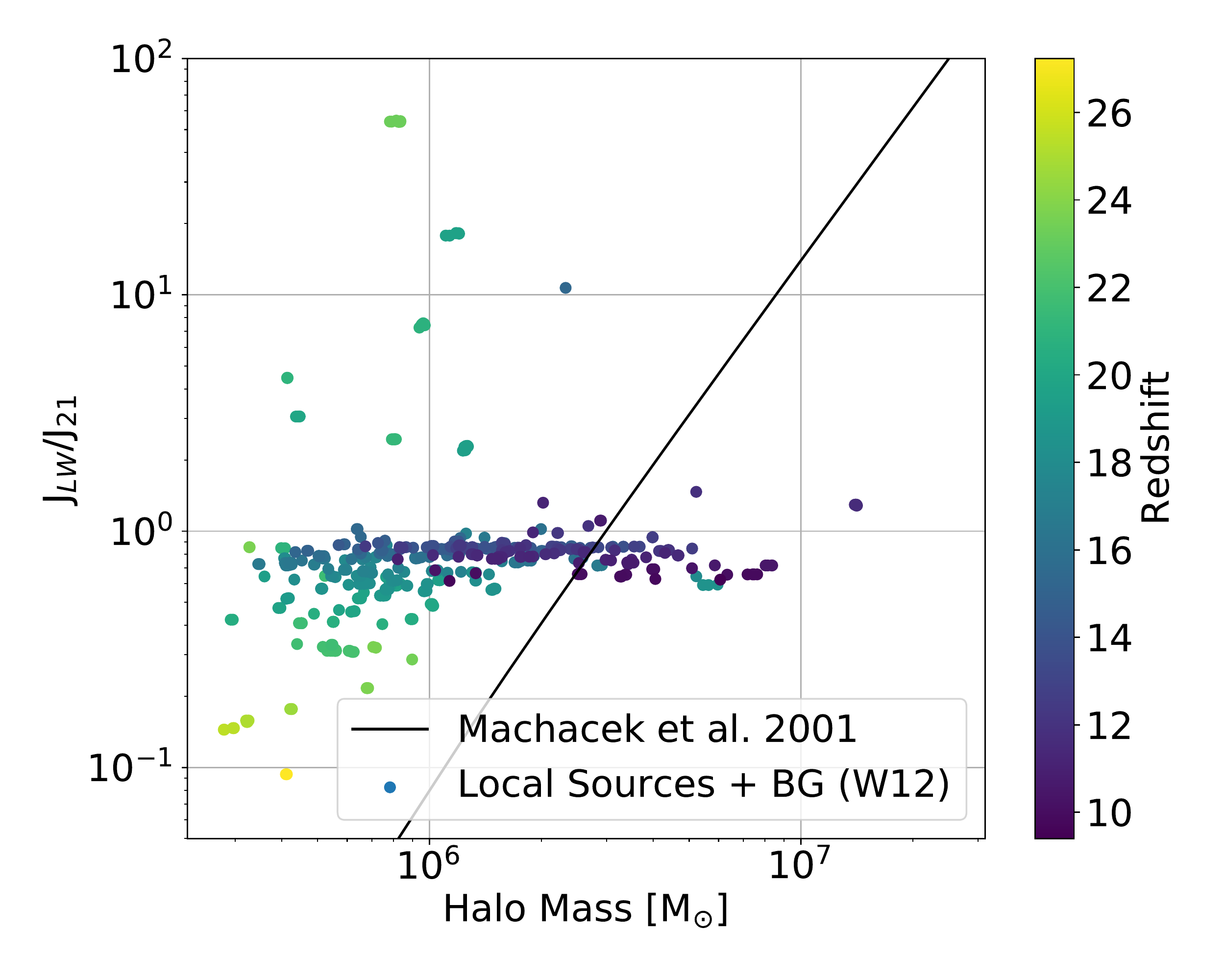}
	\caption{The average LW background for host haloes of new Pop III stars is plotted versus the host halo mass, colored by redshift. The Machacek et al. relation is plotted given the background LW in Eq. \ref{LWbg}. Almost all haloes fall below the relation, across a range of redshifts. There are extreme cases where Pop III stars form in small haloes located in a very intense LW radiation field or large haloes in relatively weak radiation fields.}
    \label{fig:jlw_mass_machacek}
\end{figure}

The LW background intensity at the instance of Pop III star formation versus the host halo mass for each Pop III halo in our dataset is shown in Figure \ref{fig:jlw_mass_machacek}. Because the LW background only varies by a factor of five as Pop III stars form, most points lie within a small intensity range, just below $1~J_{21}$. Each point is colored by the redshift where the Pop III star forms and the mass threshold from M01 (Eq. \ref{mthresh}) for the given LW background (Eq. \ref{LWbg}) is the solid black line. We see again that almost all haloes form Pop III stars below the M01 threshold. There are also a few haloes that form Pop III stars in very high $J_{\rm LW}$, below the relation. This situation arises when there are multiple Pop III stars forming at about the same time, within $\approx$ 100 pc of each other. Star formation can occur in a neighboring halo of a Pop III star whose LW radiation does not have ample time to photodissociate enough \hh{} to completely suppress star formation. The high $J_{\rm LW}$ is generally an indicator that the haloes will likely have their star formation suppressed, but in cases where Pop III star formation occurs at about the same time, star formation will not be suppressed, so long as they are a suitable distance away from each other. As there are only a few data points in this region, this situation is rare. It should be noted that there are duplicate haloes within this plot, since haloes are allowed to form multiple Pop III stars if the conditions are sufficient and each point represents an instance of Pop III formation. The grouped points in the higher end of $J_{\rm LW}$ are representative of such haloes. 

We find that a total of 84\% of the haloes forming Pop III stars lie below the M01 mass threshold over the entire simulation redshift range.  Given that the background intensity only varies by a factor of five, previous work suggested that most haloes should have formed stars around or above the M01 relationship.  However, we see a large spread in host halo masses, over an order of magnitude, and no clear relationship between the instantaneous LW intensity and the host halo mass. This behavior suggests that the LW intensity is not the primary indicator of the mass at which a metal-free halo collapses. \hh{} self-shielding is the main cause behind this conclusion because the internal dynamics and chemo-thermal evolution are now isolated from its large-scale environment and mostly depends on the local conditions within the halo. Alternatively, it is possible that there could be a correlation between the time-integrated LW background and the host halo mass. These calculations are outside the scope of this paper and require further study.

%====================================================================
\subsection{Multiple stellar systems}
%====================================================================

We now inspect the number of Pop III stars and the total mass of Pop III stars per halo. We are only tracking star-forming regions where the star particles form at densities equivalent to molecular cores that may form multiple massive stars.  Furthermore, we are not modeling the formation of low-mass stars ($<1~\Ms$) that may form out of fragmented gas at higher densities and smaller scales.  Thus our results in this section should be considered as lower limits to the number of Pop III stars.

\begin{figure*}
  \centering
  \includegraphics[width=0.48\textwidth]{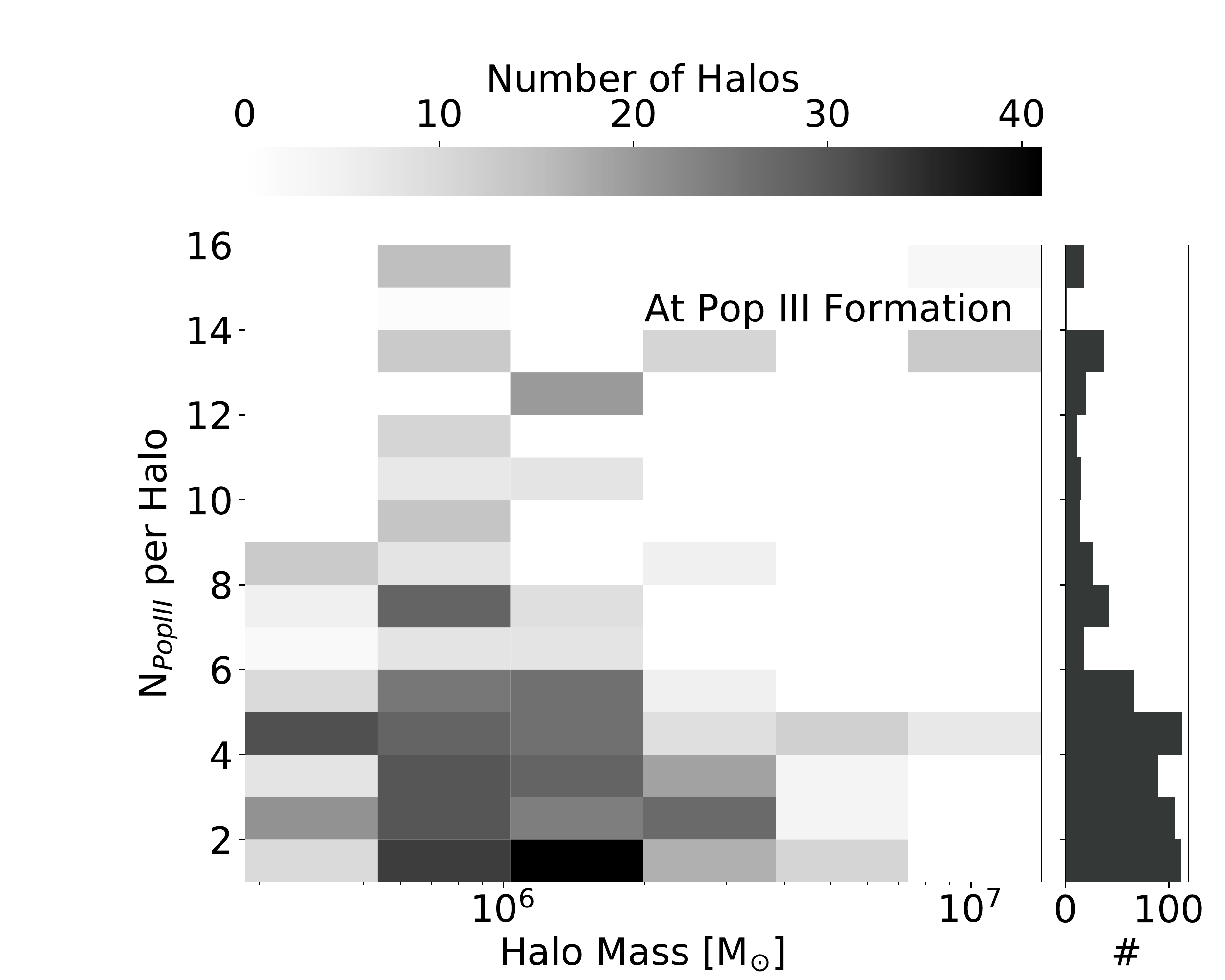}
  \hfill
  \includegraphics[width=0.48\textwidth]{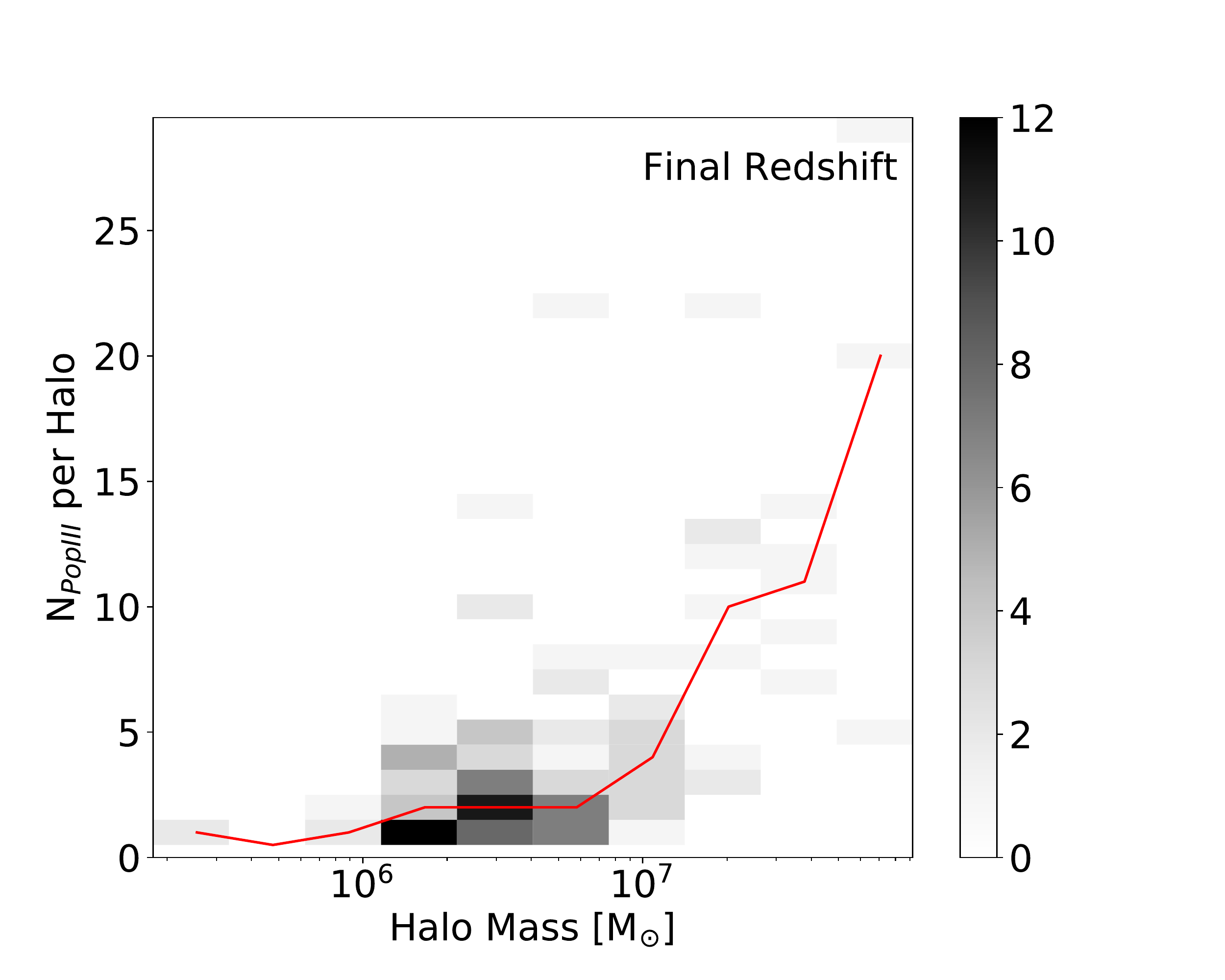}
  \caption{The number of Pop III stars per halo versus halo mass at the instance of Pop III star formation (left) and the final redshift (right). At the instance of formation, Pop III stars will form in a halo which contains a median number of four Pop III stars, with some containing as many as 16 Pop III stars. On the right, the red line indicates the median number of Pop III stars in each halo mass bin. By the end of the simulation, high mass haloes contain a large number of Pop III stars due to their long merger history with smaller haloes hosting Pop III stars.}
  \label{fig:num_p3}
\end{figure*}

\begin{figure}
	\includegraphics[width=\columnwidth]{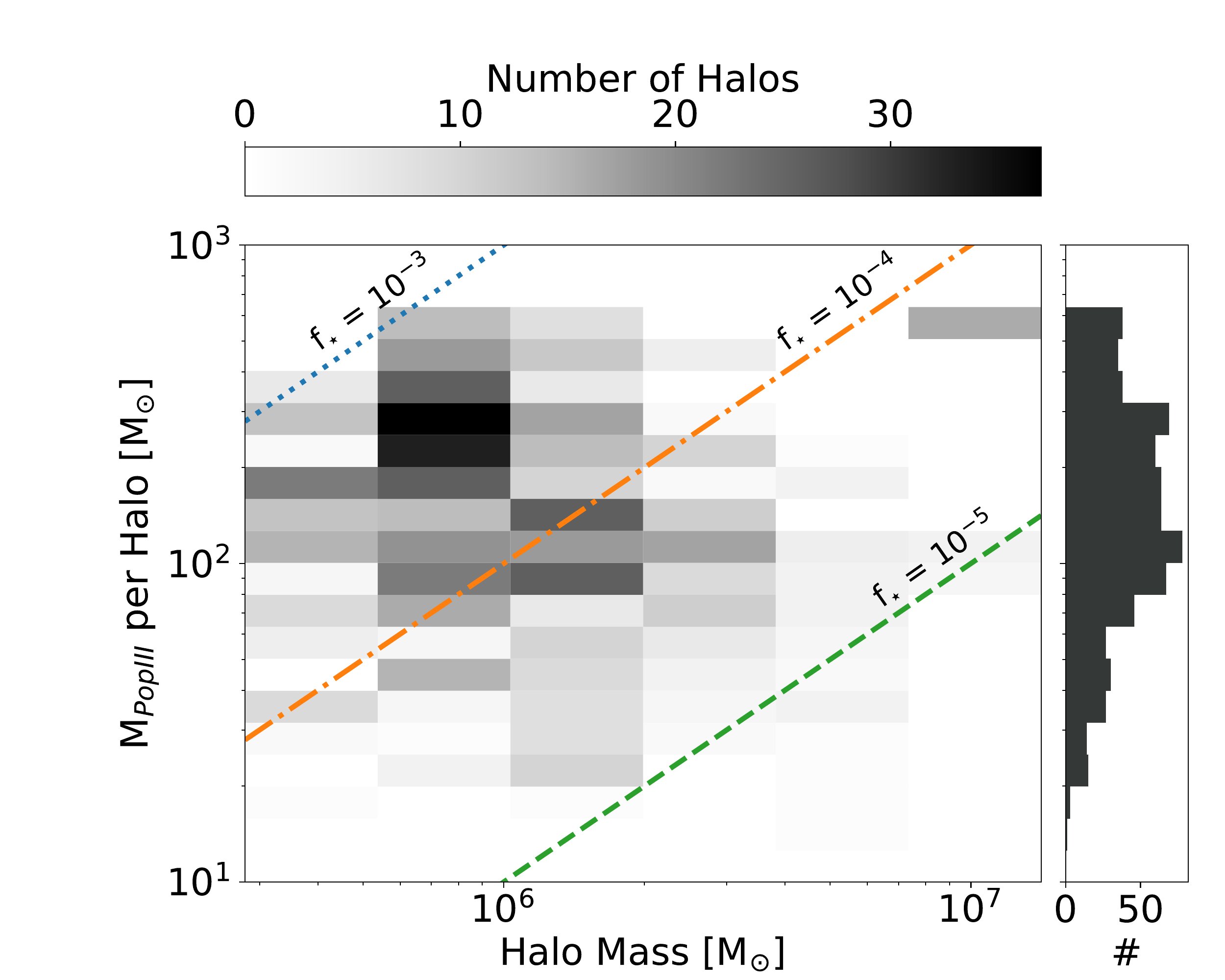}
    \caption{Total mass of Pop III stars in haloes hosting new Pop III stars versus halo mass. Lines of constant star formation efficiencies are overplotted. Most haloes form Pop III stars at efficiencies between 10$^{-3}$ and 10$^{-4}$.}
    \label{fig:totp3mass_halomass_sidehist}
\end{figure}

The left panel of Figure \ref{fig:num_p3} shows the number of Pop III stars per halo for a given halo mass, just after star formation. The histogram of the number of Pop III stars for all masses is projected on the right hand side. We find that a median number of four Pop III stars form per halo, with a maximum of 16 Pop III stars forming per halo. Since we did not restrict the number of Pop III stars that can form in a halo, we find that the conditions are often sufficient for multiple Pop III stars to form in a single halo. Out of the haloes forming Pop III stars, only 16\% of them form single Pop III stars, whereas 54\% form between two and five Pop III stars, similar to the radiation hydrodynamic simulations of \citet{Susa13, Susa14}. It should be noted that these haloes are not necessarily forming all of the Pop III stars at once. For example, a halo can form stars again if it did not host a supernova. The number of Pop III stars that form within a halo is indicative of the star formation history of that halo. Figure \ref{fig:totp3mass_halomass_sidehist} shows the total mass of Pop III stars per halo for a given halo mass, where the histogram of the total Pop III mass for all haloes is projected on the right hand side. Lines of constant star formation efficiency are overplotted. We find that most of our Pop III stars are forming between 10$^{-4}$ < f$_\star$ < 10$^{-3}$. The mean total mass of Pop III stars for all haloes is 195 M$_{\odot}$. Note that the mass of the Pop III stars depends on the chosen IMF. 

\begin{figure}
	\includegraphics[width=\columnwidth]{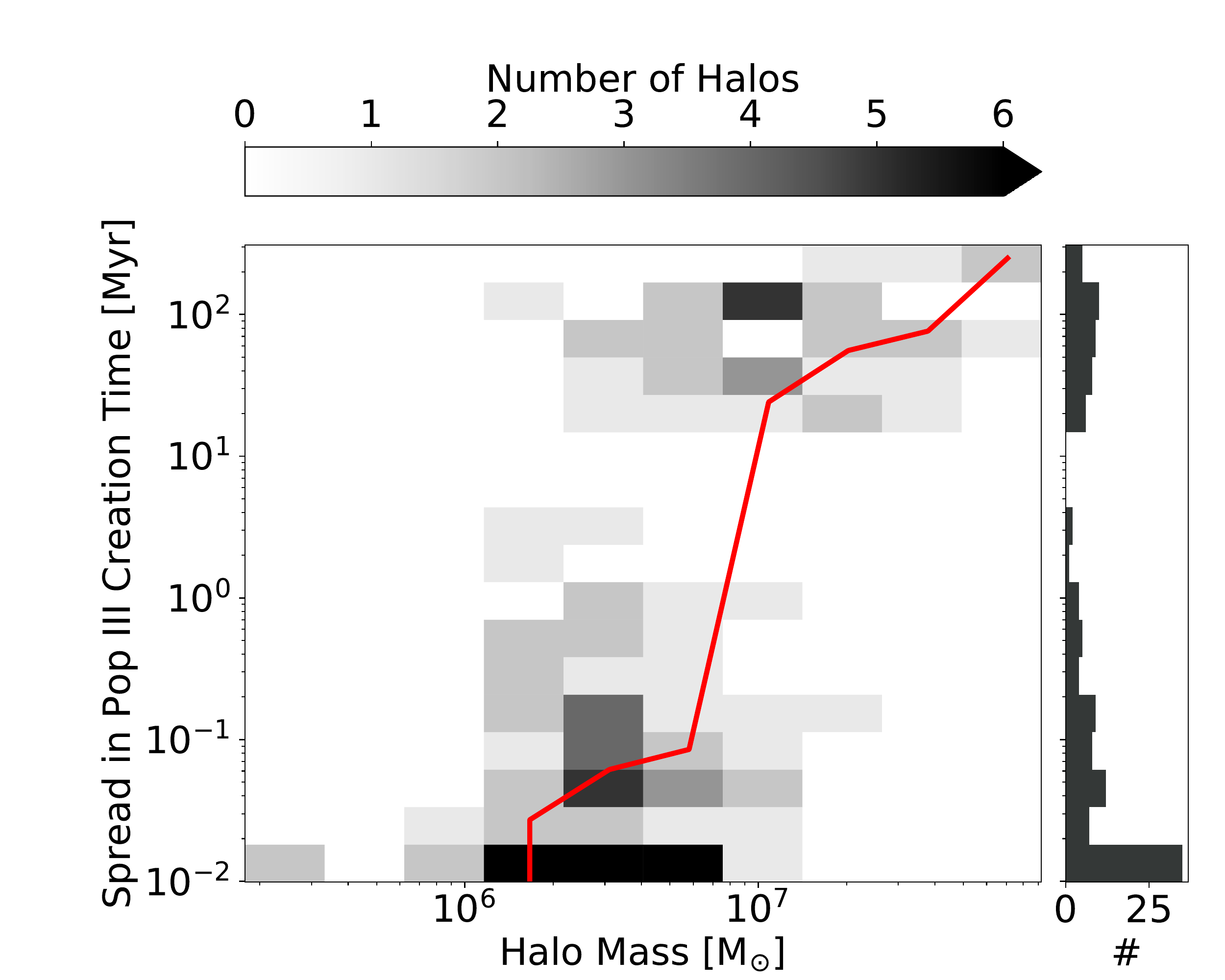}
    \caption{The spread in Pop III creation times per halo versus halo mass at the end of the simulation. The red line indicates the median spread in creation times for each halo mass bin. There are two groups that appear in this plot, haloes that have a spread of less than 10 Myr, representing haloes that form Pop III stars quickly, and haloes that have a spread of greater than 10 Myr, representing larger mass haloes that acquire old Pop III stars through mergers.}
    \label{fig:p3spread_mass}
\end{figure}

At the end of the simulation, the right panel of Figure \ref{fig:num_p3} shows the distribution of the number of Pop III stars per halo for a given halo mass, for all haloes in the simulation that host either a living or a remnant Pop III star. The red line shows the median number of Pop III stars for each halo mass bin. At low masses, the number of Pop III stars per halo sits around 2 Pop III stars, and quickly rises once the halo mass reaches $10^{7} M_{\odot}$. \citet{Xu13} found that at $z = 15$, haloes with masses at about $10^{7} M_{\odot}$ will contain between 10 and 100 Pop III stellar remnants, similar to our results. We can also look at the spread in creation times of these Pop III stars per halo, shown in Figure \ref{fig:p3spread_mass}. The haloes that have a zero spread in their Pop III creation time, meaning the Pop III stars formed at the same time, lie at $10^{-2}$ Myr. The red line indicates the median spread in creation times for each halo mass bin. There are two groups within this plot. The first group contains haloes that form their Pop III stars in a spread of < 10 Myr, with a median spread of about 0.1 Myr. The second group contains haloes that form their Pop III stars in a spread of > 10 Myr, with a median spread of about 100 Myr. The second group represents typically larger mass haloes that have acquired older Pop III stars by the end of the simulation through their long merger history. These haloes have acquired these Pop III stars from outside haloes, accounting for the large spread in Pop III creation time as well as the increased number of Pop III stars, as seen in the right panel of Figure \ref{fig:num_p3}. 

%====================================================================
\section{Discussion}
%====================================================================

%====================================================================
\subsection{Variations in the halo masses at collapse}
\label{sec:variations}
%====================================================================
Previous studies have focused on the minimum halo mass of Pop III hosts as a function of LW intensity. However, they have also found that Pop III stars often form in haloes up to an order of magnitude greater than the minimum. We also have found that Pop III star formation occurs in a similar range of host halo masses throughout our simulation.

There are three main causes of this significant variation in mass. First given our assumed IMF (Equation \ref{IMF}), 36 per cent of metal-free stars directly form a black hole without metal ejecta \citep{Heger03}, leaving the halo chemically pristine.  Second-generation stars forming later in that halo could still be metal-free if no external enrichment occurs. Thus they would form in haloes substantially larger than the minimum, as it takes tens of millions of years for the halo to recover its gas after radiative feedback \citep{Muratov13, Jeon14_Recovery}.

Second, dynamical heating from mergers and accretion provide a heating source inside the growing halo, preventing the gas from efficiently cooling via \hh{}. This turbulent stirring can delay star formation that could cool through \hh{} in an ideal situation but otherwise forms in haloes more massive than the minimum \citep{Yoshida03, Wise19}.

Finally, temporal fluctuations in the local LW radiation field can greatly affect the amount of \hh{} within a halo and thus, how efficiently the halo can cool. While this may only be a small effect in some haloes where the local LW radiation fluctuations are relatively small and star formation is distant, Pop III star formation may be significantly delayed or completely prevented in haloes that are close to active star formation sites.  In summary, the timing of Pop III star formation depends on both local -- halo histories of star formation and growth -- and environmental properties.  The combination of these three processes determines which haloes will be able to form Pop III stars, and thus results in a wide range of host halo masses.

%====================================================================
\subsection{Comparison to previous work}
%====================================================================

\begin{figure}
	\includegraphics[width=\columnwidth]{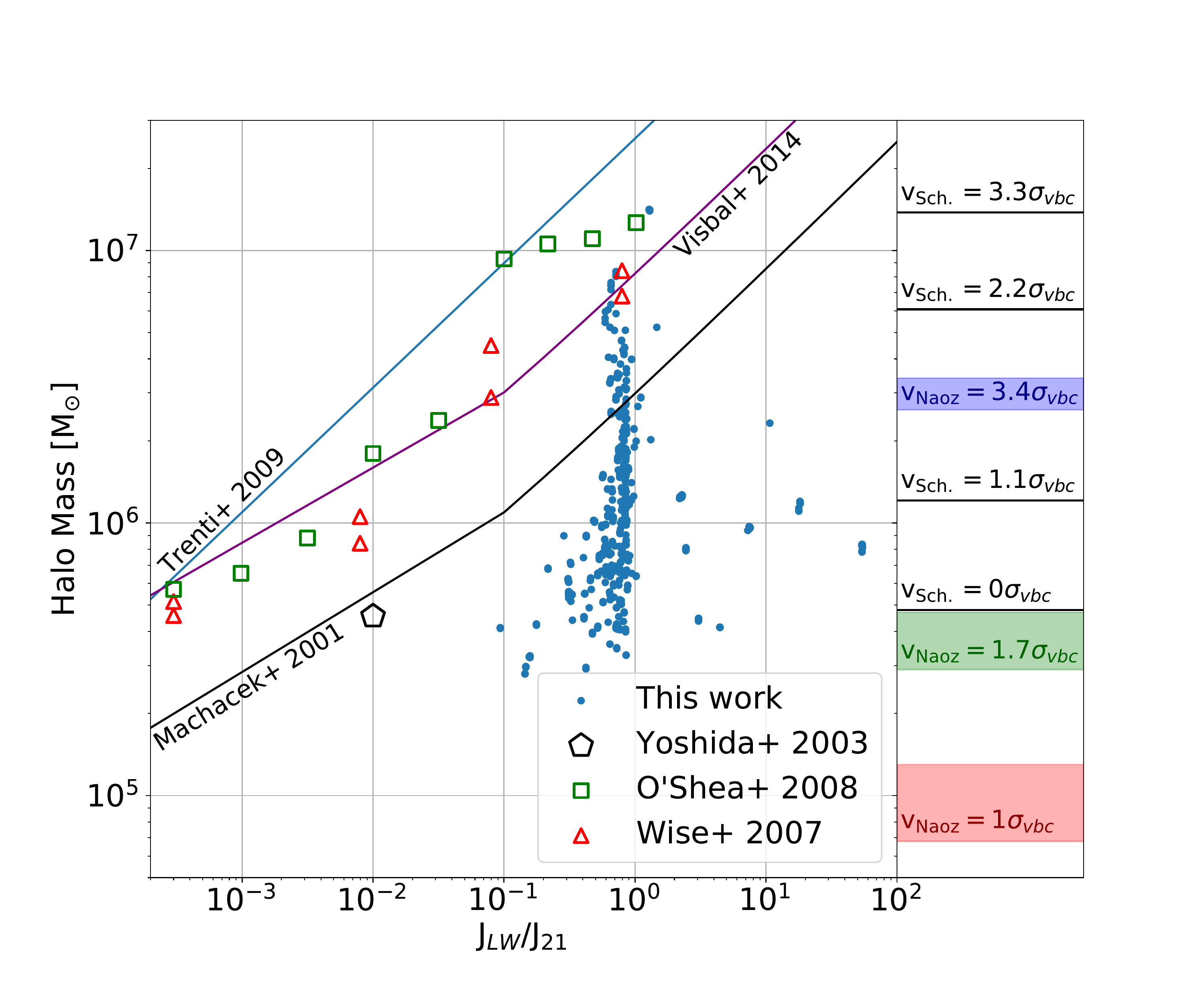}
    \caption{The halo mass versus the average LW intensity normalized by J21 versus for this work and a variety of other works. The black solid line is from M01 given our LW background, blue solid line from \citet{Trenti09_SFR} at $z=20$, the purple solid line from \citet{Visbal14} at $z=20$, blue open circle from \citet{Yoshida03}, green open squares from \citet{OShea08}, and red open triangles from \citet{Wise07_UVB}. $J_{LW}$ = 0 for the various sources is plotted here at $J_{\rm LW}$/$J_{21}$ = 10$^{-4}$. The right panel represents halo masses corresponding to different streaming velocities from \citet{Naoz13} and \citet{Schauer19}. The solid bands of halo masses correspond to the characteristic masses of haloes to contain a baryon fraction that is half of the cosmic mean for various redshifts and for the shown streaming velocities from \citet{Naoz13}. For each band, the maximum halo mass occurs at $z = 15$. For streaming velocities of $3.4 \sigma_{\rm vbc}$, $1.7 \sigma_{\rm vbc}$, and $1 \sigma_{\rm vbc}$ (5.8 \kms{} at $z=199$), the minimum halo mass occurs at $z = 17, 22, \rm{and} \ 27$, respectively. The minimum halo mass hosting cold dense gas for different streaming velocities, given at their starting redshift $z=200$, from \citet{Schauer19} are shown as horizontal black lines.}
    \label{fig:compare_JLW_mass}
\end{figure}

There has been a large amount of work done throughout the literature investigating the relationship between the LW background intensity and halo masses. \citet{Tegmark97} numerically integrated a chemical network to calculate the minimum mass a halo must have in order to cool, by considering \hh{} formation and cooling rates. They conclude that cooling by \hh{} is efficient and leads to the first formation of structures. They found a minimum halo mass needed to collapse which depends on the virial temperature of the halo and the virial redshift. At $z = 15$, this minimum halo mass is $10^{6} M_{\odot}$. \citet{Trenti09} uses a similar argument as \citet{Tegmark97}, and found that metal-free haloes can exist until z $\approx$ 6 using cosmological simulations and an analytical model for metal enrichment. They also provide some insight into whether or not Pop III supernova rates are high enough within these metal-free haloes to allow for the LSST to see these regions of the universe. They used the minimum halo mass capable of cooling via \hh{} from \citet{Trenti09_SFR} as one part of their halo mass model (see the blue solid line in Figure \ref{fig:compare_JLW_mass}, given our LW background). Neither \citet{Trenti09} nor \citet{Tegmark97} included the effects of \hh{} self-shielding, which can explain the discrepancy between our data and their mass threshold relationship (Figure \ref{fig:compare_JLW_mass}). For a LW background of $J_{\rm LW} / J_{21} = 0.2$, we see haloes hosting Pop III stars down to $4 \times 10^{5}$ M$_{\odot}$, while the relation from \citet{Trenti09_SFR} would expect to see minimum halo masses capable of cooling, and thus capable of hosting Pop III stars, at $10^7$ M$_{\odot}$, almost two orders of magnitude larger. Since we include \hh{} self-shielding, Pop III stars are allowed to form in smaller mass haloes for a given LW background. 

\citet{Visbal14} used one-zone models to study the gaseous centers of dark matter haloes and how they react to a LW background. They use a particular parametrization of the central gas density, which then follows their analytic equation for critical mass (see their Equation 4) to within a factor of two. This critical mass is plotted in our Figure \ref{fig:compare_JLW_mass} at redshift 20. Here almost all of our haloes fall below their critical mass relationship, with only the highest mass haloes rising above. \citeauthor{Visbal14} also compare their results to three-dimensional hydrodynamical simulations, where they find similar threshold halo masses as \citet{Machacek01}. \citet{Mebane18} used a semi-analytic model of star formation, including feedback properties such as a LW background, photoionization due to Pop III stars, supernovae of Pop III stars and metal-enriched stars, and chemical enrichment, to determine for how long Pop III stars will survive for and found that Pop III stars can continue to form until z $\approx$ 6. They found a minimum halo mass for hosting Pop III star formation, but with the caveat that they did not include \hh{} self-shielding. They found that when Pop III stars contribute to the LW background, the minimum halo mass for Pop III star formation is $4 \times 10^{6} M_{\odot}$ at $z = 20$, similar to results from \citet{Tegmark97}. The results from \citet{Trenti09} and \citet{Mebane18} have higher halo masses in comparison with M01, although this more closely matches results from simulations \citep[see][]{Wise07_UVB, OShea08}. 

It should be noted that because our box is fairly small, we cannot capture the cosmic variance of rare haloes and galaxies. For example, at late times, a single galaxy dominates the LW radiation field, and drives up the host halo masses. However, we would expect this to happen at different times in other cosmological volumes. Therefore, we cannot directly compare the time dependence, however, a comparison as a function of LW intensity is still valid. 

\citet{Yoshida03} used cosmological simulations to study the formation of primordial star-forming clouds. They followed the growth of structure to find where gas cools and condenses which would form the first stars, and how the effects of LW radiation may affect these gas clouds. Importantly, they included \hh{} self-shielding within haloes. In a series of simulations, they found a minimum halo mass for those haloes hosting gas clouds which may result in active star formation (see their Figure 12). They found that in the presence of a LW background of 0.01 $J_{21}$, the minimum halo mass is $7 \times 10^{5} M_{\odot}$. When \hh{} self-shielding is taken into account along with a LW background, their minimum halo mass decreases to $4.5 \times 10^{5} M_{\odot}$. This value lies close to the case where there is no LW background applied, where the minimum halo mass is $3.5 \times 10^{5} M_{\odot}$. They found that \hh{} self-shielding does appear to be an efficient mechanism enabling primordial gas cooling. In comparison with our work, we find similar minimum halo masses for each redshift, although we do see a wider range of halo masses, ranging from $2.8 \times 10^{5}$ M$_{\odot}$ to $1.4 \times 10^{7}$ M$_{\odot}$. While our haloes never experience a LW intensity as low as 0.01 $J_{21}$, we find that their minimum mass for this LW intensity lies in the same mass range as our haloes (see blue open circle in Figure \ref{fig:compare_JLW_mass}). 

\citet{Wise07_UVB} used cosmological simulations to investigate \hh{} cooling in a LW background. They found that \hh{} cooling is dominant even when there is a large LW background present. They did not include self-shielding and subsequently found halo masses at their collapse that lie well above the M01 relation (see red open triangles in Figure \ref{fig:compare_JLW_mass}). Their $J_{\rm LW} = 0$ control is plotted in Figure \ref{fig:compare_JLW_mass} at $J_{\rm LW}/J_{21} = 10^{-4}$. \citet{OShea08} used cosmological simulations to investigate Pop III star formation in various LW backgrounds. They found that due to an increased LW background, there is a delay in star formation, and thus there is an increase in the halo masses at collapse. They also ignored \hh{} self-shielding in their calculations which can again account for the increased halo masses they found compared to this work, as can be seen in Figure \ref{fig:compare_JLW_mass} (open green squares). Their halo masses are similar to the results of \citet{Wise07_UVB}. Their control of $J_{21}$ = 0 control is plotted at $J_{\rm LW}/J_{21} = 10^{-4}$. \citet{Hirano15} used cosmological simulations to study 1540 collapsing metal-free gas clouds in the early universe to derive the corresponding primordial stellar mass distribution. They find that stars form in haloes above virial masses $M_{\rm vir} = 2.1 \times 10^5 M_{\odot}$ at $z = 30$ and $M_{\rm vir} = 9.9 \times 10^5 M_{\odot}$ at $z = 10$. These masses are consistent with our results with our mean halo mass lying between these values.

%====================================================================
\subsection{Caveats}
%====================================================================

There are a few minor shortcomings to this work that should be noted. As discussed in \S2.2.1, the number of Pop III star particles that form is insensitive to merging distances in our formation algorithm down to 0.625 pc. Since this value is larger than fragmentation scales within molecular cores, it is possible that our Pop III star particles are sites for multiple Pop III star formation. With this in mind, our multiplicity results are only strengthened by the possibility of having more Pop III stars forming at these locations.

\citet{Schauer17} investigated the escape fraction of LW photons in the near and far-field and found that the LW escape fraction of atomic cooling haloes can vary significantly depending on the ionisation front within the halo. In the near-field with an ionisation front breaking out of a halo, they find that the LW escape fraction is greater than 95\%. But when the ionisation front is contained within the halo, they found that the LW escape fraction can range from 3\% to 88\%. In general, they find that the LW escape fractions in the far-field are higher than 75\%. Their work was based on previous work, done by \citet{Schauer15}, where they study LW radiation coming from single Pop III stars, rather than stellar populations. They find similar results as \citet{Schauer17}, with the exception that \hh{} shielding by neutral hydrogen decreases LW escape fractions as compared to when only \hh{} shielding is considered. In their more recent work, they find that this difference in LW escape fractions between when \hh{} shielding by neutral hydrogen is included or not is negligible.

\citet{kitayama04} investigated the transition between D-type and R-type ionization fronts within haloes to determine the escape fractions of ionizing and photodissociating photons using a one-dimensional Lagrangian hydrodynamics code. They found that in haloes with masses $< 10^6 M_{\odot}$, escape fractions are greater than 80 \%, and in larger mass haloes, with masses $> 10^7 M_{\odot}$, the escape fractions are essentially zero. Since we are not accounting for the reduced escape fraction within our haloes, we are overestimating the LW radiation coming from point sources in our simulation. We could account for this by decreasing the intensity of the stars, since the overall LW radiation would be reduced, although this should not significantly affect our results as we find little dependence on LW intensity. 

We also do not include streaming velocities between baryons and dark matter within our simulation \citep{Tselia11, Greif11_Delay, Naoz12, OLeary12, Schauer19, Hirano17_Science}. Streaming velocities can suppress star formation at very high ($z \ga 20$) redshifts only, because the streaming velocities decrease inversely with the scale factor. Pop III star formation is delayed by an average of $\delta z = 4$ by increasing the halo mass needed to overcome the bulk velocity by about a factor of three \citep{Greif11_Delay}. The streaming velocities will generally result in lower gas densities within haloes as well as an offset of the peak density from the center of the halo \citep{OLeary12}. \citet{Naoz13} estimated the minimum mass needed to retain the bulk of the baryons within the dark matter halo using a series of cosmological simulations, the results of which can be seen in Figure \ref{fig:compare_JLW_mass} (see shaded bands). These haloes only reach up to about $2 - 3 \times 10^5 M_\odot$ when the streaming velocity is about $1\sigma_{\rm vbc} = 3 \kms$ at $z=99$, which is smaller than the star forming haloes in our simulation. \citet{Naoz13} mention that at the largest streaming velocity, their simulations do not match expected values due to a small sample size of high mass haloes. 

A more recent study by \citet{Schauer19}, who used cosmological simulations to study cold gas content within haloes for varying streaming velocities, showed that high streaming velocities could restrict first star formation to atomic cooling haloes in the most extreme cases. Their minimum halo masses which host cold dense gas for various streaming velocities are shown as black horizontal lines in Figure \ref{fig:compare_JLW_mass}. \citet{Hirano17_Science} investigated the effect that streaming velocities have on massive black hole seed formation and find that in regions with high streaming velocities and for z $> 30$, first generation star formation can be suppressed in minihaloes up to the atomic cooling limit.

The results of \citet{Naoz13} and \citep{Schauer19} in Figure \ref{fig:compare_JLW_mass} show that typical streaming velocities ($\le 1\sigma_{\rm vbc} = 5.8\kms$ at $z=199$) would affect haloes below $10^6~\Ms$, especially at redshifts $z>20$ when streaming velocities are the strongest.  Our results sample minihaloes that experience such streaming velocities primarily at $z < 20$, whereas \citet{Hirano17_Science} studied much earlier structure formation. In the evolution of a halo affected by streaming velocities, it would have started to accumulate gas once above these critical masses while also being irradiated by a LW background.  Afterwards, the collapse will be controlled by the processes discussed in Section \ref{sec:variations}, and thus streaming velocities would likely not affect our main conclusion that \hh{} plays an important role in Pop III star formation.

Finally with most studies of metal-free stars, there is the caveat of its uncertain IMF that affects the individual stellar luminosities and surface temperatures of the Pop III stars.  Changes in the IMF could alter the multiplicity and total stellar mass of the system because the radiative feedback from the earlier stars can affect the thermodynamics of other molecular clouds within the same halo.  Ultimately the IMF determines the fraction of stars going supernova and the metal production in pre-galactic objects that is the root cause of suppressing and ending Pop III star formation in the early universe.

%====================================================================
\section{Conclusions}
%====================================================================
In this work, we presented the analysis of the host halo mass distribution of Pop III stars and how it relates to the LW background radiation. From our results, we conclude the following: 

\begin{itemize}
	\item We find that Pop III stars are forming in haloes with a mean mass below $10^{6}$ M$_{\odot}$, which falls well below the M01 relation, until $z=15$, at which point the mean halo mass rises above the M01 relation to a mean value of $3 \times 10^{6}$ M$_{\odot}$ due to the domination of LW radiation produced by metal-enriched stars in the most massive galaxy in the simulation box.
	\item \hh{} self-shielding allows haloes below the M01 mass relation, as small as half as previously thought, to cool and form Pop III stars.  Self-shielding creates a disconnect between the instantaneous LW background and the collapse, resulting in a broad mass distribution of metal-free star forming haloes and no strict correlation between the two quantities.
	\item Halos are likely to form multiple Pop III stars. At the instance of Pop III star formation, a halo will have formed a median number of four Pop III stars, up to a maximum of 16.
	\item By the end of the simulation, haloes with masses $>10^{7} M_{\odot}$ acquire multiple Pop III stars due to mergers with smaller mass haloes, resulting in galaxies with young and old metal-free stars. 
\end{itemize}

Our results provide another piece of the Pop III puzzle. With the inclusion of \hh{} self-shielding, we find Pop III stars forming in smaller mass haloes than previously predicted. These smaller mass haloes provide earlier and additional sites for Pop III star formation, and therefore, more metal-enrichment within haloes. We also find that Pop III stars do not necessarily have to form in isolation, in fact, they are more likely to form in a halo that has already formed multiple other Pop III stars. Since these stars are generally massive and are likely to produce stellar mass black holes, their remnants may be gravitational wave candidates detected through LIGO \citep{Hartwig16}. The exact detection signature of these gravitational waves is a topic which needs to be studied further. While we do not see any relationship between the LW intensity and Pop III host halo mass,  in order to determine exactly how the LW intensity influences the host halo mass, a more systematic approach needs to be taken. The time-integrated LW background leading up to the formation of a Pop III star would be an interesting calculation to follow up with, and may result in a correlation with the host halo mass.

The assumptions often made about the formation of Pop III stars, like how they will only form in isolation and how the LW background mainly determines the collapse mass, appear to break down when self-shielding is taken into account and restrictions about their formation are lifted. Further work needs to be done to determine why some haloes form Pop III stars and others do not, to provide insight about the formation process of these stars and thus how they will affect their surrounding pre-galactic environment. The effect of the LW intensity on host halo mass should be systematically investigated with \hh{} self-shielding to determine whether or not this background radiation really has a profound effect on Pop III formation or not. Further constraints on these primordial objects will assist multi-messenger astronomers in identifying these ancient systems.

%====================================================================
\section*{Acknowledgments}
%====================================================================

JHW is supported by National Science Foundation grants AST-1614333 and OAC-1835213, and NASA grant NNX17AG23G.  The simulation was performed on Blue
Waters operated by the National Center for Supercomputing Applications (NCSA) with PRAC allocation support by the NSF (awards ACI-1514580 and OAC-1810584). The subsequent analysis was performed with NSF's XSEDE allocation AST-120046 on the Comet resource and also on the Georgia Tech PACE compute system.  This research is part of the Blue Waters sustained-petascale computing project, which is supported by the NSF (awards OCI-0725070, ACI-1238993) and the state of Illinois. Blue Waters is a joint effort of the University of Illinois at Urbana-Champaign and its NCSA.  The freely available plotting library {\sc matplotlib} \citep{matplotlib} was used to construct numerous plots within this paper. Computations and analysis described in this work were performed using the publicly-available \enzo{} and \yt{} codes, which is the product of a collaborative effort of many independent scientists from numerous institutions around the world.

%%%%%%%%%%%%%%%%%%%%%%%%%%%%%%%%%%%%%%%%%%%%%%%%%%

%%%%%%%%%%%%%%%%%%%% REFERENCES %%%%%%%%%%%%%%%%%%

% The best way to enter references is to use BibTeX:

\bibliographystyle{mnras}
\bibliography{jwise} % if your bibtex file is called example.bib

% Alternatively you could enter them by hand, like this:
% This method is tedious and prone to error if you have lots of references
%\begin{thebibliography}{99}
%\end{thebibliography}

%%%%%%%%%%%%%%%%%%%%%%%%%%%%%%%%%%%%%%%%%%%%%%%%%%

%%%%%%%%%%%%%%%%% APPENDICES %%%%%%%%%%%%%%%%%%%%%

\appendix

%%%%%%%%%%%%%%%%%%%%%%%%%%%%%%%%%%%%%%%%%%%%%%%%%%

% Don't change these lines
\bsp	% typesetting comment
\label{lastpage}
\end{document}